\documentclass{aa}
\usepackage{graphicx, rotate}

\usepackage{txfonts}

\begin{document}

\title{
The Tully--Fisher relation at intermediate redshift
\thanks{Based on observations with the European Southern
Observatory Very Large Telescope (ESO-VLT), 
observing run IDs 65.O-0049, 66.A-0547 and 68.A-0013.} 
\fnmsep\thanks{Our full data table 
is available in 
electronic form via anonymous ftp to \texttt{cdsarc.u-\-strasbg.fr}
}}

\author{
A. B\"ohm\inst{1}\and 
B. L. Ziegler\inst{1}\and 
R. P. Saglia\inst{2}\and       
R. Bender\inst{2}\and
K. J. Fricke\inst{1}\and \\
A. Gabasch\inst{2}\and 
J. Heidt\inst{3}\and 
D. Mehlert\inst{3}\and 
S. Noll\inst{3}\and 
S. Seitz\inst{2}
}

\offprints{A. B\"ohm, \\ e-mail: boehm@uni-sw.gwdg.de}

\institute{Universit\"atssternwarte G\"ottingen, Geismarlandstra{\ss}e 11,
	   37083
	   G\"ottingen, Germany
\and Universit\"atssternwarte M\"unchen, Scheinerstra{\ss}e 1,
	    81679 M\"unchen, Germany
\and Landessternwarte Heidelberg, K\"onigstuhl, 69117 Heidelberg, Germany
}

\date{Received 2 September 2003 / Accepted 26 January 2004}

\abstract{
Using the Very Large Telescope in Multi Object Spectroscopy mode,
we have observed a sample of 113 field spiral galaxies in the 
FORS Deep Field (FDF) with redshifts in the range $0.1<z<1.0$. 
The galaxies were selected based on apparent  brightness ($R<23^m$)
and encompass all late spectrophotometric types from Sa to Sdm/Im.
Spatially resolved rotation curves have been extracted for 77 galaxies and 
fitted with synthetic velocity fields taking into account all observational 
effects from inclination and slit misalignment to seeing and slit width.
We also compared different shapes for the intrinsic rotation curve.
To obtain robust values of $V_{\rm max}$,
our analysis is focused on galaxies with rotation curves that    
extend well into the region of constant rotation velocity at large radii.
If the slope of the local Tully--Fisher relation (TFR) is held fixed,
we  find evidence for a mass--dependent luminosity evolution which is as 
large as up to $\Delta M_B \approx -2^m$ for the lowest--mass
galaxies, but is small or even negligible for the highest--mass systems 
in our sample. In effect, the TFR slope is shallower at $z \approx 0.5$ 
in comparison to the local sample. We argue for a mass--dependent evolution 
of the mass--to--light ratio. An additional population of blue, low-mass 
spirals does not seem a very appealing explanation.
The flatter tilt we find for the distant TFR is in contradiction to the 
predictions of recent semi--analytic simulations.

}   

\titlerunning{The Tully--Fisher relation at intermediate redshift}
\authorrunning{A. B\"ohm et al.}

\maketitle


\section{\label{intro}Introduction}

Ever since the relation between the luminosity $L$ and the maximum rotation 
velocity $V_{\rm max}$ of spiral galaxies was first observed 
(Tully \& Fisher \cite{TF77}), the physical origin of its slope and
scatter, as well as the possible evolution thereof over different cosmic epochs have
been subject to debate both in theoretical and observational studies.
Within the last few years, the Tully--Fisher relation (TFR) has been
put into the framework of a Fundamental Plane (FP) for spiral galaxies that
introduces the disk scale length $r_{\rm d}$ as a third parameter
(e.g.  Burstein et al. \cite{Bu97}).
Similiar to the FP of dynamically hot galaxies, i.e. stellar systems that are
stabilized due to random motion (e.g.  Dressler et al. \cite{Dre87}),
the  
spiral FP has smaller scatter edge--on than its two--dimensional projections.
This may be understood in terms of two dominant parameters for disk galaxies,
for example the mass and angular momentum (Koda, Sofue \& Wada \cite{KSW00}).

Numerical simulations within hierarchical Cold Dark Matter (CDM)-dominated 
cosmologies have been successfully used to reproduce the observed slope of the local
TFR, whereas the numerical zero points were offset due to dark halos with too high 
concentrations (e.g.  Navarro \& Steinmetz \cite{NS00}). 
The TFR slope is predicted to remain constant with cosmic look--back time 
in such N-body simulations; nevertheless the modelling of realistic stellar
populations at sufficient resolution remains a challenge. 

Other theoretical approaches focus more on the chemo--spectrophotometric
aspects of disk galaxy evolution. 
For example, Boissier \& Prantzos (\cite{BP01}) used the ``hybrid'' approach
(Jimenez et al. \cite{Ji98}) that relates the disk surface density to the
properties of the associated DM Halo, and calibrated it to reproduce the 
observed colors of local spirals. Compared to these, the authors predict higher
luminosities for large disks and lower luminosites for small disks at
redshifts $z>0.4$.
A similiar evolution is found by Ferreras \& Silk (\cite{FS01}).
By modelling the mass--dependent chemical enrichment history 
of disk galaxies with the local TFR as a constraint, the authors find a 
TFR slope that increases with  look--back time 
(i.e., for a parameterisation $L \propto V_{\rm max}^\alpha$, $\alpha$ increases
with redshift).
 
In the last decade, many observational studies of the local TFR 
have produced very large samples with $N_{\rm obj} \approx 1000$ 
(e.g.  Haynes et al. \cite{Hay99}), not only to derive 
the slope and scatter with high accuracy, but also to map the peculiar 
velocity field out to $cz \approx 15000$\,km\,s$^{-1}$ 
(e.g.  Mathewson \& Ford \cite{MF96}). 
Other groups used spirals, partly with cepheid--calibrated distances, to 
measure the Hubble constant.
For example, Sakai et al. (\cite{Sa00}) derived a value of 
$H_0 = (71~\pm~4)$\,km\,s$^{-1}$ with this method.

At higher redshifts, robust measurements of rotation velocities 
and luminosities become increasingly difficult. 
This is partly because of the low
apparent magnitudes of the galaxies, but also due to the limited intrinsic 
spatial resolution (see Sect.~\ref{vmax} for a more detailed description
of this effect).
A number of samples with 10-20 objects in the regime
$0.25< \langle z \rangle <0.5$ have been observed in recent years 
to estimate a possible evolution in luminosity by comparison to the local TFR. 
The results of these studies were quite discrepant:
e.g. Vogt et al. (\cite{Vog96,Vog97}) find only a modest increase in luminosity
of  $\Delta M_B \approx -0.5^m$, whereas Simard \& Pritchet (\cite{SP98}) and
Rix et al. (\cite{Rix97}) derive a much stronger brightening with
$\Delta M_B \approx -2.0^m$. A study of 19 field spirals by
Milvang-Jensen et al. (\cite{Mil03}) reveals a value of
$\Delta M_B \approx -0.5^m$ and shows evidence for an increase with
redshift. Another sample of 19 spirals by Barden et al. (\cite{Ba03}) which
covers the high redshifts $0.6<z<1.5$ yields a value of 
$\Delta M_B \approx -1.1^m$. 

It seems likely that some of these results are affected by 
the selection criteria. For example, Rix et al. selected blue colors with
$(B-R)_{\rm obs}<1.2^m$, Simard \& Pritchet strong [O\,{\small II}] 
emission with equivalent widths $>$20\AA, while Vogt et al. partly chose
large disks with $r_{\rm d}>3$\,kpc. 
The two former criteria prefer late--type spirals, whereas the latter
criterion leads to the overrepresentation of large, early--type spirals.
Additionally, due to the small samples, all these studies had to assume that 
the local TFR slope holds valid at intermediate redshift.
We will further discuss this issue in Sect.~\ref{discuss}. 

Based on a larger data set from the DEEP Groth Strip Survey 
(Koo \cite{Ko01})
with $N \approx 100$ spirals in the range 
$0.2~<~z~<~1.3$, Vogt (\cite{Vog01}) finds a constant TFR slope and a 
negligible rest--frame $B$-band brightening of less than 0.2\,mag.
In a more recent publication from this group which investigates the
luminosity--metallicity relation, an evolution both in slope and zero
point is observed (Kobulnicky et al. \cite{Ko03}), in the sense
that the luminosity offsets are largest at the low--luminosity end
of the sample and smallest at the high--luminosity end. 
The authors argue that low--luminosity galaxies
could have either undergone a decrease in luminosity 
or an increase in the metallicity in the last $\sim8$\,Gyrs. 

Preliminary results from our TF project indicating a mass--dependent 
luminosity evolution of distant field spirals have been presented in a
letter (Ziegler et al. \cite{Zie02}). In this paper, we will describe
the derivation of the maximum rotation velocities, the galaxies' 
structural parameters and the luminosities in  more detail and present 
the data table of the enlarged, full sample. 
Complementary to our approach in Ziegler et al., the
complete analysis will be restricted here to galaxies with rotation curves 
that extend well into the region of constant rotation velocity at large radii,
i.e. spirals that yield robust values of $V_{\rm max}$. In addition,
different shapes for the intrinsic rotation curves will be compared.
Finally, we will discuss potential environmental effects on the sample
and present the galaxies' virial masses.
 
The paper is organized as follows. In Sect.~\ref{sample}, we describe
our selection procedure and the observations. Data reduction will be outlined
in Sect.~\ref{reduction}. The extraction of rotation curves and $V_{\rm max}$
derivations are described in Sect.~\ref{vmax}, followed by the details
of the transformations from apparent to absolute magnitudes
in Sect.~\ref{mags} and the presentation of the data table in
Sect.~\ref{data}. We will then construct the distant TFR 
in Sect.~\ref{tfr} and discuss the results in Sect.~\ref{discuss}.
A summary is given in Sect.~\ref{conclude}.

Throughout this article,
we will assume the concordance cosmology with
$\Omega_{\rm m}$ = 0.3, $\Omega_\Lambda$ = 0.7 and 
$H_0$ = 70\,km\,s$^{-1}$\,Mpc$^{-1}$.


\section{\label{sample}Sample selection and observations}

Our sample consists of galaxies in the 
FORS Deep Field (FDF), a sky region near the south galactic pole  
with deep $UBgRIJK$ photometry 
and visible completeness limits similiar to the Hubble Deep Fields.
For a description of the field selection criteria
and the total $N \approx 8750$ object catalogue of the FDF, 
we refer to Heidt et al. (\cite{Hei03}). 

The basis for spectroscopy target pre--selection was the
FDF photometric redshifts catalogue (Bender et al. \cite{Ben01}).
To keep the selection function as simple as possible, 
the only spectrophotometric costraints were that the galaxies should
have a spectral energy distribution (SED) later than E/S0 and 
an apparent total brightness of $R\le23^m$. 
This limit was chosen to gain  S/N~$\approx$~5 in  emission lines
at intermediate resolution within 2-3 hours integration time with the 
Very Large Telescope (VLT).
To ensure the visibility of either the [O\,{\small II}]\,3727 doublet,
H$\beta$ or [O\,{\small III}]\,5007 within the wavelength range of the
600R grism of FORS, the upper limit for the photometric
redshift was $z_{\rm phot}\le 1.2$. Since the basic aim was to derive
spatially resolved rotation curves of the galaxies, (most of the) objects with 
inclinations $i<40^\circ$ were rejected. 

In Multiobject Spectroscopy (MOS) mode, FORS offers 19 individually
moveable slits. The setups were prepared with the FIMS 
(FORS Instrument Mask Simulator) package.
According to the position angles, we subdivided the target galaxy sample into 
bins of 30$^\circ$ to minimize geometric distortions in the final
rotation curves. Thus, a set of 6 MOS masks was necessary to cover all
orientations. In the case of some objects, either the inclination limit of 
40$^\circ$, the limit of 15$^\circ$
deviation between position angle and slit orientation, or the magnitude
limit had to be exceeded to fill all slitlets.

The first observations were carried out with FORS2 mounted on 
VLT Unit Telescope 2 (UT2) 
in September and October 2000. Each mask was exposed for
3 $\times$ 3000\,s.
The slit widths were set to 
one arcsecond and grism 600R with order separation filter GG435 was used,  
yielding a spectral resolution of ${\rm R} \approx 1200$.
FORS was operated in its standard resolution configuration of 0.2\arcsec/pixel.
The seeing conditions covered the range $0.43\arcsec~\le~{\rm FWHM}~\le~0.81\arcsec$
according to the Differential Image Motion Monitor (DIMM). In October 2001,
an additional set of three MOS masks was observed with FORS1 mounted
on VLT UT3 with the same instrument configuration as in 2000 with seeing of
0.74\arcsec~$\le$~FWHM~$\le$~0.89\arcsec. In total, spectra of 129 spirals out of
156 candidates were taken (the latter number includes targets which slightly 
exceed the brightness limit).


\section{\label{reduction}Data reduction}

The reduction procedures were implemented in the ESO-MIDAS 
(Munich Image Data Analysis System) environment.
For all three observing runs, the bias sets taken at the beginning and end
of the night showed a very stable two--dimensional structure, so all bias 
frames from the nights of the 2000 and 2001 observations were used to 
generate two master biases.
The individual bias frames have been normalized to the same median count rate
and were median--averaged, followed by a slight Gaussian filter smoothing
($\sigma_{x,y}=3$\,pixel).
The overscan region of the science frames was used to determine the offset
constant for final master bias subtraction. Since the dark current was 
very uniform across the CCDs, an extra dark subtraction has been neglected.

Spectroscopy flatfield (FF) through--mask exposures with FORS were taken 
by default with two sets of 
lamps switched on alternately. Since either the 
upper or lower half of the frames were not usable due to contaminating light
from the gaps between the slits, the appropriate regions from both sets  of 
FF frames were extracted and re--combined after multiplicative 
normalization and median--averaging. 
The slit regions of calibration and science frames were then
individually extracted.
The FF slit exposures were approximated by a polynomial fit of sixth degree in
dispersion direction (X axis) to account for the CCD response curve, 
which was used for normalization prior to the correction of the pixel--to--pixel
variations.  

\begin{figure}[t]
\resizebox{\hsize}{!}{\includegraphics{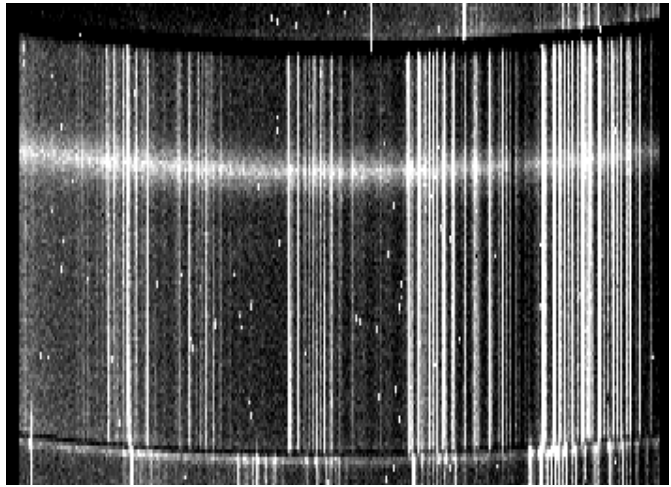}}
\caption{\label{mos}
Upper part of a raw FORS1 MOS frame with 3000 seconds exposure time showing the
region of slit \#2. Wavelength is increasing parallel to the X axis from left
to right.
A relatively bright ($R=19.94^m$) galaxy spectrum is overlaying the
night sky emission lines. 
Note that the magnification scale is larger along the Y axis to 
demonstrate the distortions by the focal reducer in this direction.}
\end{figure}

In the next step, the geometric distortions caused by the focal reducer were
corrected. For the slits at the bottom and top of the CCD, where the 
distortions were maximum, the curvature of spectral features corresponded to
a displacement of up to 5-6 pixel both in X and Y direction 
(see Fig.~\ref{mos} for an illustration).
A polynomial fit of second degree was fitted to the galaxy spectra in each slit 
to derive the curvature along the spatial axis. Based on this fit,
the science and wavelength calibration spectra were rectified
with an accuracy of 0.1 pixel (corresponding to 0.02\arcsec).  
Tests revealed that the flux conservation of this procedure was accurate to
within a few percent. The distortions along the X axis could be
corrected during standard wavelength calibration. In the calibration
exposures, the HgCd lamp was switched on additionally to the He, Ar and 
Ne lamps in order to gain a sufficient number of emission lines below 
$\sim$\,5800\,\AA.  
For the two--dimensional dispersion relation, polynomial fits
of third and first degree were used in the direction of the X and Y axes, 
respectively. The typical r.m.s. of the relation was 0.03-0.04\,\AA\ at a
stepsize of 1.08\,\AA\ per pixel.

The night sky emission was fitted column by column with
first order fits, unless a galaxy spectrum was located at the 
extreme edges of the slit. In those cases, zero order fits yielded the best
results. Night sky subtraction was performed individually for each mask exposure. 
Prior to the final addition of the three exposures, the optical center of each
galaxy along the Y axis has been determined by fitting a Gaussian. 
If necessary, the spectra were shifted by integer values (two pixels at maximum
in a few cases)
to ensure consistent profile centers to within at least half a pixel,
corresponding to 0.1\arcsec.
A weighted addition was used if the average seeing varied by more than 
$\sim$\,20 percent between the exposures.


\section{\label{vmax} Rotation velocity derivation}

\subsection{\label{zs}Redshift distribution}

Out of the 129 galaxies of which spectra were taken, redshifts could be determined 
for 113 spirals, including three cases of secondary objects which were covered
by a MOS slit by chance. 
The 16  galaxies without spectroscopic redshifts populate the extreme faint 
end of our 
apparent brightness distribution with a median of $\langle R \rangle = 22.9^m$.
Since only one of these spirals features colors of a very 
late--type SED,
the S/N of the remainders may be just too low to yield detectable emission lines.
According to the photometric redshifts, 
the [O\,{\small II}]\,3727 doublet is possibly redshifted out of the
wavelength range of the R600 grism in the case of 8 targets,
while for 5 other galaxies in the regime $0.2<z_{\rm phot}<0.5$,
only [O\,{\small III}] or H$\beta$ emission 
is potentielly covered, which is weaker than H$\alpha$ or [O\,{\small II}]
for typical spirals.

\begin{figure}[t]
\resizebox{\hsize}{!}{\rotate[r]{\includegraphics[bb=39 27 566 768]{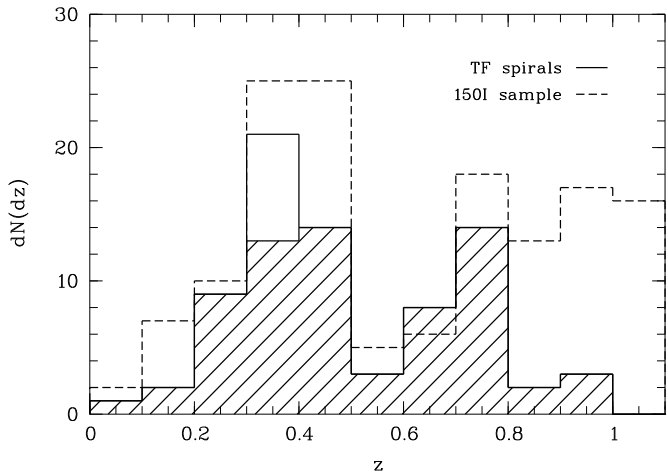}}}
\caption{\label{zdistr}
Redshift distribution of the 77 FDF spirals with usable rotation curves
(thick solid line). 
The shaded region gives the distribution after exclusion of 8 possible members
of a cluster at $z=0.33$. 
The dashed line represents the lower resolution spectroscopy sample
observed with grism 150I as part of the FDF high--z study (Noll et al. \cite{No03}).
Both data sets feature a redshift ``gap'' around $z\approx$\,0.6,
see text for details.}
\end{figure}

Fig.~\ref{zdistr} shows the redshift distribution of our sample,
restricted to objects with appropriate rotation curves for
the TF analysis, see Sect.~\ref{vfmod} for the constraints.
The median redshift is $\langle z\rangle=0.45$.
It is likely that the bimodal 
shape is a combination of the two
following effects.

Firstly, as has been outlined in Heidt et al. (\cite{Hei03}), the 
southwestern corner of the FDF
covers the outskirts of a galaxy cluster at $z=0.33$.
Allowing a spread in redshift of $\Delta z=0.01$, we find that a maximum of
8 galaxies in our sample could be members of this cluster.
The small redshift ``bump'' at $z\approx0.3$ can be attributed to these
objects.

Secondly, the sample contains relatively few galaxies around redshift 
$z\approx0.6$.
For comparison, we also show the distribution of 
144 galaxies at $z<1.1$ from
the FDF high--$z$ campaign (Mehlert et al. \cite{Me02}, Noll et al.
\cite{No03}) which are not contained in our sample. 
The high--$z$ spectra were taken with the same instrument configuration
as the TF data, except that the low--resolution grism 150I was used
which covers a much broader wavelength range of $\sim$\,6000 \AA\ in
observer's frame. 
The large number of galaxies with $z\le1$ which did not enter
our TF survey can be attributed to the much fainter brightness limits of the
high--$z$ study ($I\le24.5^m$), the inclusion of elliptical galaxies 
and the lack of constraints to the inclinations. 
It is clear from Fig.~\ref{zdistr} that both samples
feature the same redshift ``gap''.

As a test, we verified our selection criteria on the most recent version of 
the FDF photometric redshifts catalogue, finding that only a handfull of types 
later than Sa could have been missed by the original selection.
We therefore conclude that the volume probed with the FDF probably
contains fewer galaxies at $z\approx0.6$
than the neighbouring redshift bins and that the distribution of the TF 
spirals is unlikely to reflect a selection bias or an observational effect.

\subsection{\label{class}Spectrophotometric classification}

To gain a classification of our spectra, three criteria were used. 
The SED model parameter from the
photometric redshifts catalogue, which is related to the 
star formation e--folding time of the fitted templates, can be transformed
into the de Vaucouleurs scheme (ranging from $T=1$ for Sa to $T=10$ for Im). 
This is useful especially for spectra with very few identifiable lines 
and/or low S/N, but has the disadvantage that dust reddening 
might induce a classification of too early a type for highly inclined spirals.

We therefore also performed a comparison to the SED templates in the
Kennicutt catalogue (\cite{Ken92a}) with a focus on the relative line
strengths of the H$\beta$/[O\,{\small III}] emission.
As a third indicator, the rest--frame
[O\,{\small II}]\,3727 equivalent widths were derived and correlated to type
with the values for local spirals given by Kennicutt (\cite{Ken92b}) as a reference. 

A combination of these three criteria yielded 14 spirals of
type $T\le3$ in the redshift range $0.11~\le~z~\le~0.89$ with a median of 
$\langle z \rangle=0.42$,
43 galaxies with $T=5$ covering $0.09~\le~z~\le~0.97$ with $\langle z \rangle=0.45$ 
and 20 objects with $T\ge8$ in the regime $0.23~\le~z~\le~0.97$ with
$\langle z \rangle=0.51$.

\subsection{\label{rcextr}Rotation curve extraction}

The measurements of the rotation velocity as a function of radius were
performed in a semi--automatic manner. 
In the first step, 100 columns of the spectrum centered on the considered
emission line were averaged to get a profile of line plus continuum along the 
spatial axis. This profile was approximated with a Gaussian to derive the optical
center to within 0.1\arcsec. For a few objects located very close to the 
slit edge, the center had to be redefined manually. Then, the emission
lines were fitted row by row. 
To enhance the S/N, three neighbouring rows were averaged prior to the
emission line fitting; for very weak lines, this ``boxcar'' was enlarged to 
five rows (corresponding to one arcsec).
In the case of the [O\,{\small II}]\,3727 doublet, two Gaussians with equal 
FWHM and an observer's frame separation of $2.75\,(1+z)$\,[\AA] were assumed as a 
line profile approximation, while a  single Gaussian was used
for [O\,{\small III}]\,5007, $H\beta$ or $H\alpha$, 
with the latter being visible only in four spectra.
The red- and blueshifts along the spectral axis due to rotation 
were measured relative to the observed 
wavelength of the line at the optical center and converted into velocity
shifts after cosmological correction by a factor of $(1+z)^{-1}$.
This position--velocity information defines an {\it observed}\, 
rotation curve (RC).

\begin{figure*}[t]
\centering
\vspace*{0.75cm}
\resizebox{14cm}{!}{\includegraphics[bb=28 90 550 770]{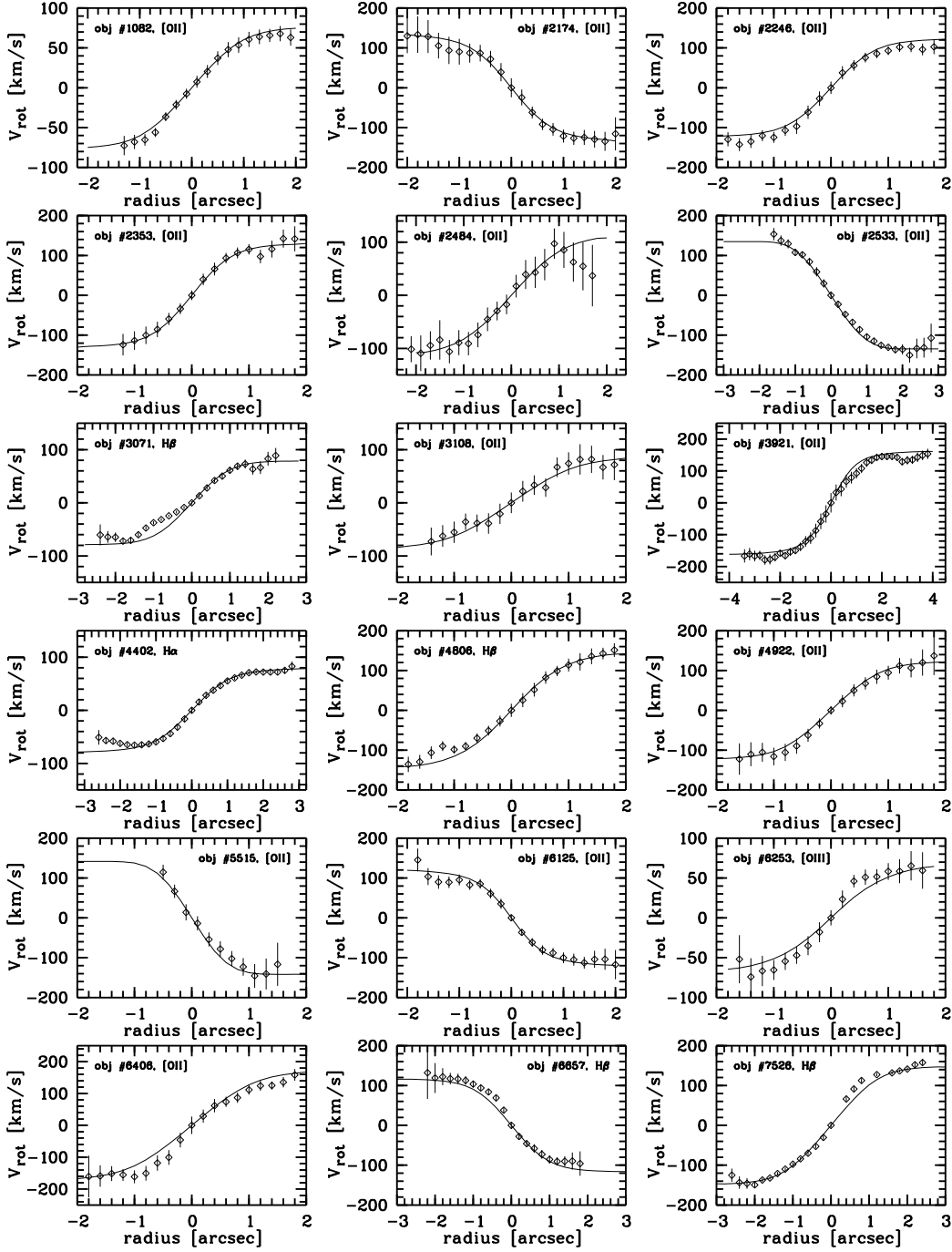}}
\caption{\label{rcall}
Examples of 
rotation curves from our data. 
Observed rotation velocity is plotted against the projected distance from
galaxy center. Error bars denote the errors of the Gaussian fits to the emission
lines. Object numbers and emission lines used are given for each spiral 
(either [O\,{\scriptsize II}]\,3727, [O\,{\scriptsize III}]\,5007,
H$\alpha$ or H$\beta$).
The observed rotation curves have been modelled with synthetic velocity fields 
(solid lines) which take into account
geometric distortions due to inclination and slit misalignment as well
as seeing and optical beam smearing, see Sect.~\ref{vfmod} for
details. The small spatial extent of the curve of galaxy \#5515 to the left 
hand side is due to contaminating light from a foreground elliptical. 
}
\end{figure*}

Each curve was visually inspected prior to the RC modelling.
Seven objects had too low a S/N to derive spatially resolved rotation velocities.
We also rejected RCs with strong asymmetries or other signatures
of substantial kinematic disturbances, ``solid--body'' rotators 
and objects that did not show any rotation at all within the
measurement errors.
In total, the RCs of 77 spirals were appropriate for the $V_{\rm max}$
derivation.
We present a range of examples in Fig.~\ref{rcall}.

\subsection{\label{vfmod}Rotation curve modelling}

For distant, apparently small galaxies, the effect of the slit width
on the observed rotation velocity as a function of radius, 
$V_{\rm rot}^{\rm obs}(r)$,
must be considered.
At redshift $z=0.5$, a scale length 
of 3\,kpc --- typical for an 
$L^{\ast}$ spiral --- corresponds
to approx. 0.5\,arcsec only, which is half the slit width used in our MOS
observations. 
Any value of $V_{\rm rot}^{\rm obs}(r)$ is therefore an integration 
perpendicular to the spatial axis (slit direction), a phenomenon which 
is the optical equivalent to ``beam smearing'' in radio observations. 
If not taken into account, this effect could lead to an underestimation of the
intrinsic  rotation velocities.

We overcame this problem by generating {\it synthetic} RCs.
Our approach is similiar to the procedure described by Simard \& Pritchet
(\cite{SP99}) but differs from their fitting method in some ways. 
E.g., the scale length of the emitting gas
is a fixed parameter in our algorithm (see below), and the deviation angle
between apparent disk major axis and slit direction is taken into account.   
Moreover, in our approach we do not fit simulated 2D spectra to observed 
2D spectra but simulated RCs to observed RCs.

For the simulation, one has to assume an {\it intrinsic}\, rotational law
$V_{\rm rot}^{\rm int}(r)$. 
As a first of three variants, we used a simple shape
with a linear rise of $V_{\rm rot}^{\rm int}$ at small radii, turning over into
a region of constant rotation velocity where the Dark Matter Halo dominates the
mass distribution. 
This can be achieved with the parameterisation
\begin{equation}
\label{young}V_{\rm rot}^{\rm int}(r) = \frac{V_{\rm max}\,r}{(r^a + r^a_0)^{1/a}}
\end{equation} 
(e.g.  Courteau \cite{Cou97}) with a factor $a$ that tunes the
sharpness of the ``turnover'' at radius $r=r_{\rm 0}$, and a constant 
rotation $V_{\rm rot}^{\rm int}(r) = V_{\rm max}$ for $r \gg r_{\rm 0}$. 
We used a range of values for $a$ on a set of $\sim$\,20 RCs
and found that $a=5$ best reproduced the observed shape of the turnover
region. To minimize the number of free parameters, we kept $a$ fixed
to that value for all objects in the further analysis.
However, due to the heavy blurring of the curves, $a$ is the
least critical parameter in Eq.~\ref{young}.


The turnover radius $r_{\rm 0}$ was assumed to be equal to the scale
length of the emitting gas, which is larger than the scale length 
$r_{\rm d}$ derived from continuum emission. This is discussed in
Ryder \& Dopita (\cite{RD94}) and Dopita \& Ryder (\cite{DR94})
using observational and theoretical approaches, respectively.
Since the structural parameters like the continuum scale lengths were 
measured in an $I$-band VLT image
(see Sect.~\ref{mags1}), we derived the corresponding gas scale length via
\begin{equation}
\label{rdgas}r_0=(2-z/2)\,r_{\rm d}. 
\end{equation}
This equation
yields $r_0 \approx 2\,r_{\rm d}$ for the least distant FDF spirals
(for which $I$ corresponds to rest--frame $R$)
and $r_0 = 1.5\,r_{\rm d}$ at $z=1$, 
where $I$ corresponds to rest--frame $B$. In other words, Eq.~\ref{rdgas}
is used to gain a correlation between rest--frame scale length and
gas scale length that is in compliance with the results of
Ryder \& Dopita.
We also tested this equation directly on a few high S/N emission line profiles 
in our spectra. 
It should be noted however, that the $V_{\rm max}$ derivation 
is much less sensitive to 
the scale length than to the inclination and the misalignment angle. 
This is mainly due to the smoothing by the instrumental PSF.

We alternatively used two other templates
of intrinsic RC shapes within the Universal Rotation Curve (URC) framework 
of Persic \& Salucci (\cite{PS91}, cited as the URC91 hereafter) 
and Persic et al. (\cite{PS96}, URC96 hereafter).
The authors introduced a dependence of the RC morphology on luminosity:
Only $L^\ast$ spirals have a constant $V_{\rm rot}$ at large radii,
whereas the curves of sub--$L^\ast$ galaxies are rising even beyond
a characteristic radius, and very luminous objects have a negative
gradient at large radii. 
The characteristic radii which define a characteristic rotation velocity
that can be used for the TF analysis differ slightly between the two
approaches. In the URC91 form, this radius is equal to 2.2 scale
lengths, whereas it is as large as 3.2\,$r_{\rm d}$ for the URC96.
Throughout the following sections, the quantity $V_{\rm max}$ will refer
to the usage of equation \ref{young}, whereas $V_{2.2}$ and $V_{\rm opt}$
denote an input of the URC91 and URC96, respectively.

\begin{figure}[t]
\resizebox{\hsize}{!}{\rotate[r]{\includegraphics[bb=46 36 560 560]{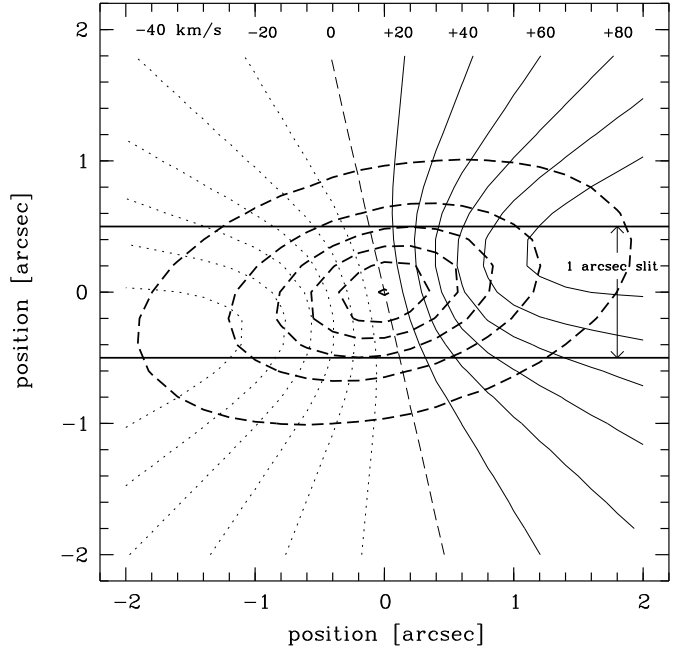}}}
\caption{\label{vf}
Simulated velocity field for galaxy \#4922 
inclined by $i$\,=\,64$^\circ$ 
with an intrinsic rotation curve of $V_{\rm max}$\,=\,156\,km/s. 
Thin solid and dotted lines indicate iso--velocity zones ranging from $+120$\,km/s
to $-120$\,km/s rotation in line--of--sight, i.e. comprising a factor $\sin i$.
The thick solid lines denote the slit,
thick dashed lines sketch the isophotes of the galaxy, with the outermost 
corresponding to 2$\sigma$ sky noise level in the 3000\,s reference
$I$-band image (see Sect.~\ref{mags1}).
The misalignment between slit and apparent major axis is $\delta$\,=\,13$^\circ$.}
\end{figure}

In the next step of the simulation, the two--dimensional velocity field was 
generated on the basis of the respective intrinsic rotational law 
(e.g.  Eq.~\ref{young} \& \ref{rdgas}),
tilted and rotated according to the observed inclination and position angle 
of the respective galaxy. The field was then weighted by
the normalized surface brightness profile,
i.e. brighter regions contribute stronger to the velocity 
shift at a given radius. 
Following this, the field was convolved. The Point Spread Function
was assumed to be Gaussian with a FWHM determined from the mean DIMM values
during the spectroscopy. These values had to be slightly increased depending 
on the
redshifted wavelength of the emission line 
from which the observed RC was derived.
We determined the correlations for this increase by measuring the FWHM on sets of 
VLT exposures in $B$, $R$, $I$, and comparing these to the according DIMM
values.
After the double folding, a strip of one arcsecond width was extracted from the 
velocity field at an angle that matched the slit misalignment in the observation 
(see Fig.~\ref{vf}). 
The final computation step was an integration perpendicular to the simulated slit,
i.e. the projection of the velocity field strip onto the spatial axis.
In all of the following, we will refer to this resulting position--velocity
model as the {\it synthetic rotation curve}, 
whereas the rotational law (e.g.  Eq.~\ref{young})
that is used as  input for the modelling procedure
will be referred to by the term {\it intrinsic rotation curve}.
The {\it observed} RC as described in Sect.~\ref{rcextr} would 
directly reproduce the intrinsic RC only for a disk of 90$^\circ$ inclination
(i.e., perfectly edge--on)
that is observed with an infinitely thin slit at infinitely large spatial and
spectral resolution.

\begin{figure}[t]
\resizebox{\hsize}{!}{\includegraphics{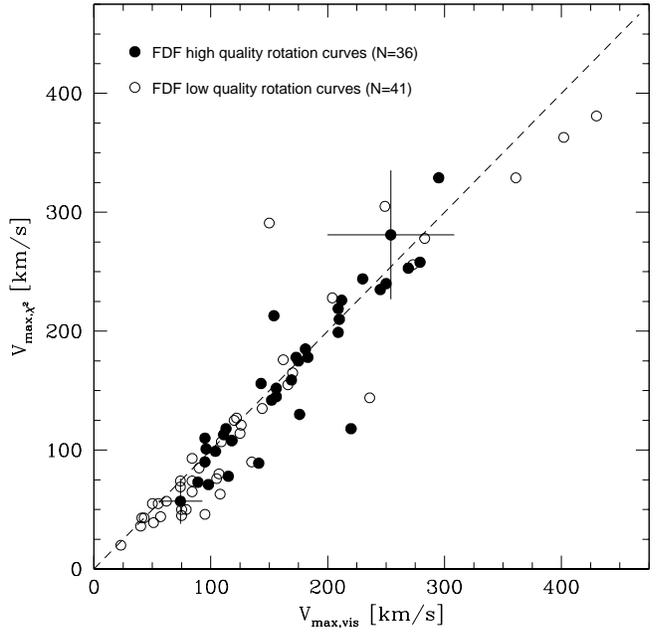}}
\caption{\label{vmax1}
Comparison between the $V_{\rm max}$ derivation by visual alignment of
observed and synthetic rotation curves (x axis) and via  $\chi^2$-fits 
(y axis).
Typical error bars are shown for two objects.
}
\end{figure}

In our approach, the intrinsic quantity $V_{\rm max}$ is the only free 
parameter that tunes the reproduction of an observed RC by a synthetic curve.
Only in the case of 3 spirals, the gas scale length had to
be kept as a second free parameter, since the values based on Eq.~\ref{rdgas}
were too large.
For the complete data set, we derived $V_{\rm max}$ by a visual 
comparison of synthetic and observed RC and, alternatively, via a  
$\chi^2$-fitting procedure based on the errors from the RC extraction.
These two methods are compared in Fig.~\ref{vmax1}. The sample of 77 spirals
is subdivided according  data quality: Curves which clearly probe the
region of constant rotation velocity at large radii are considered high quality
($N_{\rm obj}=36$),
whereas RCs with smaller extent or asymmetries are included in
the low quality sub--sample ($N_{\rm obj}=41$).  

The error on $V_{\rm max}$ is assumed to be 
\begin{equation}
\label{verr}\sigma_{V_{\rm max}}^2 = \sigma_{\chi^2}^2 
+ V_{\rm max}^2\, (\tan i)^{-2}\,  \sigma_i^2
+ V_{\rm max}^2\,  (\tan \delta)^{2}\,  \sigma_\delta^2 
\end{equation}
Here, the first term on the right hand side is the error from the $\chi^2$-fits 
of the synthetic to the observed RCs,
covering the range 3\,km/s $\le \sigma_ {\chi^2}\le$ 59\,km/s. 
The last two terms are the propagated
errors of the respective uncertainties of the inclination and the misalignment 
angle. To derive the contributions of the errors $\sigma_i$ and 
$\sigma_\delta$, we used a simple geometric correlation between the observed and
intrinsic rotation velocity,
\begin{equation}
V^{\rm int}_{\rm rot} = V^{\rm obs}_{\rm rot}\, (\sin i)^{-1}\, (\cos \delta)^{-1}.
\end{equation}
For the high quality sample, the absolute and relative errors on $V_{\rm max}$ 
fall in the respective ranges 3\,km/s~$\le~\sigma_{V_{\rm max}}~\le$~135\,km/s
and $0.03\le \sigma_ {V_{\rm max}}/V_{\rm max} \le0.61$, with a median
$\langle \sigma_ {V_{\rm max}}/V_{\rm max} \rangle~=~0.19$.

As can be deduced from  Fig.~\ref{vmax1}, the by--eye ``fits'' and the
$\chi^2$-fits are consistent within the
errors for the majority of the objects. Nevertheless, a systematic trend towards
low values of $V_{\rm max,\chi^2}$ with respect to the visually derived
$V_{\rm max,vis}$ is evident for slow rotators. An inspection of the
observed RCs revealed that the discrepancies mainly arise in cases of 
asymmetric shapes. The $\chi^2$-fits also are
weighted towards the inner parts of the curves by the higher S/N of the emission 
lines at smaller radii. 
This is a disadvantage since the outer parts of an observed RC
are the most robust source of $V_{\rm max}$ in the modelling procedure.  
Moreover, we found no correlation between the reduced $\chi^2$ values
and our definition of RC quality, except for curves of perfect 
symmetry.
For these reasons, we used the visually derived values for the 
TF analysis. A further discussion of this topic will follow in Sect.~\ref{urcs}.

In cases of multiple usable emission features in a spectrum, the RCs 
based on different lines were mostly consistent within the errors.
For these objects, $V_{\rm max}$ was derived from the curve with the
largest covered radius and highest S/N. 

\begin{figure}[t]
\resizebox{\hsize}{!}{\includegraphics{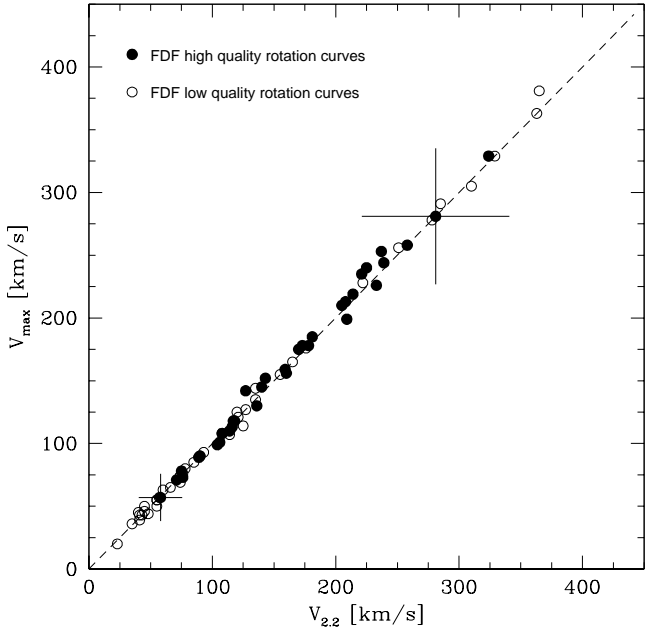}}
\caption{\label{vmax2}
Comparison between rotation curve fitting with a simple
``rise--turnover--flat'' shape (y axis) and the universal rotation curve 
shape as introduced by Persic \& Salucci (\cite{PS91}, x axis).
Solid symbols denote high quality curves which cover the region of constant 
rotation  velocity at large radii.
Typical error bars are shown for two objects.
}
\end{figure}

\begin{figure}[t]
\resizebox{\hsize}{!}{\includegraphics{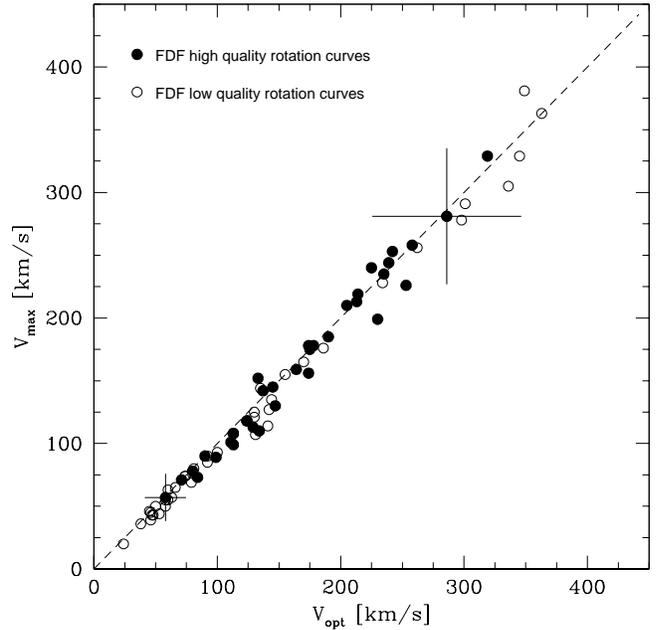}}
\caption{\label{vmax3}
Comparison between rotation curve fitting with a simple
``rise--turnover--flat'' shape (y axis) and the universal rotation curve 
shape as introduced by Persic et al. (\cite{PS96}, x axis).
High quality curves (solid symbols) cover the region of constant rotation 
velocity at large radii. 
Typical error bars are shown for two objects.
}
\end{figure}

We show a consistency check of the three alternatives for the intrinsic
RC shape in Figs.~\ref{vmax2} and \ref{vmax3}. The results using the simple
``rise--turnover--flat'' shape via Eq.~\ref{young} are in
agreement with the URC91  to within 5\% for
79\% of the FDF spirals and to within 10\% for 95\% of the galaxies,
without a detectable dependence on the absolute values.
The URC96 yields rotation velocities at $R_{\rm opt}\approx3.2\,r_{\rm d}$
which on the mean are larger by 7\% than $V_{\rm max}$, with slightly increasing
differences towards slow rotation.
This partly is an effect of the different characteristic radii in the
two parameterisations of the universal rotation curve, which correspond
to 2.2\,$r_{\rm d}$ for the URC91 and $3.2\,r_{\rm d}$ 
for the URC96, respectively. 
However, the slight differences between $V_{\rm max}$ and $V_{\rm opt}$
do not affect the results from our TF analysis, see
Sect.~\ref{urcs}.

\section{\label{mags}Photometry}

\subsection{\label{mags1}Luminosity profiles}

Since the FDF imaging had been done under varying seeing
conditions (ranging, e.g., from 0.46\arcsec\ to 0.89\arcsec\ FWHM in the $I$-band,
see Heidt et al. \cite{Hei03} for a description of the image stacking),
only a limited number of images was used for the measurements of the structural
parameters. 
To combine high spatial resolution with
sufficient S/N at the outer isophotes of the TF objects, the 10 $I$-band 
frames with the best seeing were co--added. This yielded a reference image 
with 3000\,s total exposure time, 0.49\arcsec\ FWHM and a 50\% completeness
limit of $\sim$\,25.1$^m$. 

The disk light distributions were fitted with exponential profiles.
A bulge component could not be accounted for at the given resolution.
We considered slight variations of the PSF across the reference frame
with six stars in the range $18<I<20$.  
Our algorithm minimizes $\chi^2$ in the parameter space span by
inclination, position angle, scale length and central flux. 
The sample galaxies cover a range of $25^\circ \le i \le 80^\circ$
with a median of $\langle i \rangle=53^\circ$ and a median error of 
$\langle \sigma_i \rangle=5^\circ$.

As a test of the accuracy of our ground--based luminosity profile analysis, 
we also performed measurements on
a ``VLT simulation'' of the Hubble Deep Field North 
(HDF-N, Williams et al. \cite{Wi96}). 
The original drizzled images
were re--binned to a scale of 0.2\arcsec\ per pixel and convolved with
a Gaussian PSF of 0.49\arcsec\ FWHM to match the characteristics of our 
reference $I$-band frame. We selected 40 objects with a
variety of $B/T$ ratios and $i>30^\circ$ from the electronically 
available data published by Marleau \& Simard (\cite{MS98}). 
The authors did apply two--component fits to more than 
500 galaxies in the HDF-N
using the GIM2D package 
(Galaxy Image Two--Dimensional, Simard et al. \cite{Sim02}).
We found that for $B/T<0.4$, the mean difference 
between the inclinations 
derived from the ``VLT simulation'' frame and the original GIM2D values 
was $\Delta i = 1.8^\circ \pm 7.7^\circ$. 

The median inclinations of our TF sample sub--divided according to SED type
are $\langle i \rangle=51^\circ$ for $T\le3$, 
$\langle i \rangle=53^\circ$ for $T=5$ and 
$\langle i \rangle=51^\circ$ for $T=8$, i.e. potentially
unresolved bulge components of early--type spirals 
in our ground--based imaging did not introduce a detectable bias.

\subsection{\label{mags2}Rest--frame magnitudes}

Total apparent magnitudes were derived with the Source Extractor package
(Bertin \& Arnouts \cite{BA96}). This program offers different
algorithms for the photometry. We used the
{\it Mag\_auto} routine with variable elliptical apertures which is
based on the ``first moment'' algorithm by Kron (\cite{Kro80}) since 
it best reproduces the total magnitudes of extended sources. 

We will now briefly discuss the issue of the $k$-correction.
If available broad band information is limited to a few filters
(in most previous studies two or even only one), the different wavelength ranges 
covered by a passband in observed frame and rest--frame introduce a 
strong dependence of the possible $k$-correction accuracy on SED type.
At redshift $z=0.6$, e.g., the difference between $T=4$  and $T=6$
corresponds to a change of half a magnitude in the transformation from
$B_{\rm obs}$ to $B_{\rm rest}$, $k_B$
(e.g.  Frei \& Gunn \cite{FG94}). 
Values of $k_B$ differ even more for earlier types,
thus even a slight misclassification can introduce a substantial offset in the
derived luminosity if observations are limited to one or two filters. 
In contrast to this, our TF project greatly benefits from the multi--band imaging 
of the FDF: The photometry in $B$, $g$, $R$ and $I$ enables us to use
the filter that best matches the rest--frame $B$-band to transform an apparent 
magnitude $X$ into absolute magnitudes $M_B$ up to the highest redshifts in the sample.

We computed the $k_B$-correction for our objects via synthetic photometry
on SEDs in the range $1 \le T \le 10$. 
As templates, we used the spectra published by
M\"oller et al. (\cite{Moe01}) which were generated with evolutionary
synthesis models. 
The SEDs were redshifted by re--calculating the original flux $F_0$ 
at wavelength $\lambda$ according to
\begin{equation}
\label{flux1}F(\lambda,z)=\frac{F_0[\lambda/(1+z)]}{(1+z)}
\end{equation}
(e.g.  Contardo, Steinmetz \& Fritze-v. Alvensleben \cite{CSF98}).
Transformation from the apparent magnitude of a spectrum of 
type $T$ at redshift $z$ observed with a FORS filter $X$ to 
the un--redshifted spectrum in Johnson $B$ yields $k_B(X,T,z)$. 
Respective filters used for the input magnitudes were
$B_{\rm FORS}$ for $z<0.25$, $g_{\rm FORS}$ for $0.25 \le z < 0.55$,
$R_{\rm FORS}$ for $0.55 \le z < 0.85$ and $I_{\rm FORS}$ for $z \ge 0.85$.
This $k$-correction is much less sensitive to spectral type than a transformation
$B_{\rm obs} \rightarrow B_{\rm rest}$, especially at higher redshifts:
e.g., $k_B(R,1,0.6)-k_B(R,5,0.6)=0.05^m$ whereas 
$k_B(B,1,0.6)-k_B(B,5,0.6)=0.64^m$. 

For testing purposes, we additionally derived the colors of the templates purely
within the Johnson-Cousins Filter system and compared them to the values
published by Fukugita et al. (\cite{Fuk95}), finding typical absolute
deviations of $0.03^m~\le~\vert\Delta(X~-~Y)\vert~\le~0.08^m$. 

The second critical correction that has to be applied to the 
observed magnitudes is the inclination--dependent intrinsic absorption $A_B^i$
by the dust disks of the objects. Like, e.g., 
Vogt et al. (\cite{Vog96}, \cite{Vog97}) and Milvang-Jensen et al. (\cite{Mil03}),
we adopted the approach by Tully \& Fouqu\'e (\cite{TF85}). 
It is based on geometric assumptions and usable for inclinations up to 
$i=80^\circ$, i.e. for our complete sample.
The dust disk scale height is assumed to be half the scale height of the
luminous disk, relative dust content is independent of mass or type,
with an optical depth of $\tau=0.55$.
For the objects in our sample, $A_B^i$ spans values from 
0.30$^m$ at $i=25^\circ$ to 0.96$^m$ at $i=80^\circ$. We emphasize that the 
absorption for face--on disks is finite in this convention: 
$A_B^i=0.27^m$ for $i=0^\circ$. 
Our multi--band photometry ensures to compute the absorption 
for the same rest--frame wavelength interval at all the covered redshifts.

For the galactic absorption at the coordinates of the FDF, we adopted the values 
which are given in Heidt et al. (\cite{Hei03}), ranging from $A_B^g=0.076^m$ to 
$A_I^g=0.035^m$. Let $D_\Lambda(z,H_0)$ be the distance 
modulus (e.g.  Peebles \cite{Pe93}) in the concordance cosmology, then
the transformation from total apparent magnitude $m_X$ to absolute 
magnitude $M_B$ is given by
\begin{equation}
M_B = m_X - D_\Lambda(z,H_0) - k_B(X,T,z) - A_B^i - A_X^g 
\end{equation}

If one assumes an SED classification error of $\Delta T=2$
(i.e., an Sc spiral could be misclassified as an Sb and vice versa), 
corresponding to an
uncertainty of the $k$-correction of $\sigma_k~\le~0.1^m$ for all covered types and 
redshifts,
the errors in absolute magnitude become
\begin{equation}
\label{errabs}\sigma_{M_B}^2 = \sigma_X^2+\sigma_k^2+\sigma_{A_B^i}^2,
\end{equation}
where $\sigma_X$ is the random photometric error in the respective filter
and $\sigma_{A_B^i}$ the uncertainty of the intrinsic absorption correction 
via error propagation from the inclination error.  
Thanks to the very deep imaging, the random 
photometric errors are 
only $0.01^m$ on average and $0.03^m$ at maximum for the FDF spirals. 
Based on a comparison of our calibrations 
with archived FORS zero points, we estimate the
systematic photometric errors to be $\le0.01^m$; these are neglected.
The total errors $\sigma_{M_B}$ of our complete sample fall into the range 
$0.07^m\le\sigma_{M_B}\le0.21^m$.


\section{\label{data}The data table}

We present the data of the spirals from our sample in Table~\ref{datatab}.
This table is available in electronic form via anonymous ftp to
\texttt{cdsarc.u-\-strasbg.fr}.
The respective columns have the following meaning: 

ID --- Original entry number from the FORS Deep Field photometric catalogue,
see Heidt et al. (\cite{Hei03}) for, e.g., the J2000 positions or 
aperture magnitudes in $U$, $B$, $g$, $R$, $I$, $J$ and $K$.

\begin{table*}[p]
\centering
\caption[]{\label{datatab}The data of the 77 spirals from the FORS Deep Field 
Tully--Fisher sample. 
}
\small
\begin{tabular}{cccccccccccccrc}
\hline \\
ID & \hspace{-0.2cm} $i$ & $\delta$ & $r_{\rm d}$ & $z$ & $T$ & $X$ & $k_B$ & \hspace{-0.2cm}$A^g_X$ & \hspace{-0.2cm}$A^i_B$ & 
\hspace{-0.2cm}$D_\Lambda$ & $M_B$  & $V_{\rm max}\hspace{6.5pt} $ & \hspace{-0.2cm}$B$$-$$R$ & note \\ 
 & \hspace{-0.2cm} [deg] &  [deg] & [\ $''$] & & & [mag] & [mag] & \hspace{-0.2cm}[mag]  & \hspace{-0.2cm}[mag] & \hspace{-0.2cm}[mag] & 
[mag] & [km/s]\hspace{0.7pt} & \hspace{-0.2cm}[mag] & \\  \\
\hline \\
 400 & 60 & 23 & 0.70 & 0.4483 & 10 & 22.63 & $+$0.25 & 0.06 & 0.49 & 41.98 & $-$20.15$\pm$0.13 &  74$\pm$19  & \hspace{-0.2cm}0.21 & L \\  
 745 & 42 & 61 & 0.49 & 0.6986 &  5 & 21.06 & $-$0.49 & 0.04 & 0.35 & 43.14 & $-$21.98$\pm$0.08 & 402$\pm$167 & \hspace{-0.2cm}1.00 & L \\  
 870 & 53 &  8 & 0.53 & 0.2775 &  5 & 20.77 & $-$0.07 & 0.06 & 0.42 & 40.76 & $-$20.40$\pm$0.08 &  96$\pm$6   & \hspace{-0.2cm}0.73 & H  \\
1082 & 49 & 10 & 0.50 & 0.4482 &  8 & 22.14 & $+$0.36 & 0.06 & 0.39 & 41.98 & $-$20.66$\pm$0.08 & 118$\pm$16  & \hspace{-0.2cm}0.75 & H \\
1224 & 49 &  5 & 0.34 & 0.3989 & 10 & 23.26 & $+$0.16 & 0.06 & 0.39 & 41.68 & $-$19.03$\pm$0.13 &  43$\pm$16  & \hspace{-0.2cm}0.26 & L \\
1327 & 31 &  1 & 0.62 & 0.3141 &  3 & 20.50 & $+$0.11 & 0.06 & 0.31 & 41.07 & $-$21.05$\pm$0.07 & 204$\pm$36  & \hspace{-0.2cm}0.91 & L \\
1449 & 36 & 30 & 0.66 & 0.1140 &  5 & 20.52 & $+$0.40 & 0.08 & 0.33 & 38.62 & $-$18.90$\pm$0.08 &  51$\pm$16  & \hspace{-0.2cm}0.67 & L \\
1476 & 77 & 10 & 0.80 & 0.4360 &  3 & 22.72 & $+$0.47 & 0.06 & 0.85 & 41.91 & $-$20.58$\pm$0.09 & 283$\pm$26  & \hspace{-0.2cm}1.52 & L \\
1569 & 37 &  3 & 0.34 & 0.4625 &  8 & 24.16 & $+$0.39 & 0.06 & 0.33 & 42.06 & $-$18.69$\pm$0.13 &  57$\pm$50  & \hspace{-0.2cm}0.69 & L \\
1625 & 57 & 10 & 0.38 & 0.2304 &  5 & 23.37 & $+$0.74 & 0.08 & 0.46 & 40.30 & $-$18.20$\pm$0.09 & 122$\pm$42  & \hspace{-0.2cm}0.84 & L \\
1655 & 51 &  0 & 0.49 & 0.3377 &  5 & 21.88 & $+$0.12 & 0.06 & 0.41 & 41.25 & $-$19.96$\pm$0.08 & 126$\pm$8   & \hspace{-0.2cm}0.84 & L  \\
1699 & 48 & 11 & 0.21 & 0.2299 &  8 & 22.67 & $+$0.68 & 0.08 & 0.38 & 40.30 & $-$18.77$\pm$0.13 &  55$\pm$15  & \hspace{-0.2cm}0.11 & L \\
1834 & 49 &  7 & 0.25 & 0.3475 &  5 & 24.06 & $+$0.15 & 0.06 & 0.39 & 41.33 & $-$17.88$\pm$0.13 &  62$\pm$37  & \hspace{-0.2cm}1.39 & L \\
1928 & 60 & 12 & 0.29 & 0.7179 & 10 & 22.89 & $-$0.46 & 0.04 & 0.49 & 43.22 & $-$20.40$\pm$0.13 & 105$\pm$29  & \hspace{-0.2cm}0.42 & L \\
2007 & 62 &  5 & 0.40 & 0.7175 & 10 & 22.68 & $-$0.46 & 0.04 & 0.52 & 43.21 & $-$20.63$\pm$0.11 &  84$\pm$20  & \hspace{-0.2cm}0.48 & L \\
2067 & 35 & 19 & 0.40 & 0.7942 &  5 & 21.87 & $-$0.32 & 0.04 & 0.32 & 43.48 & $-$21.65$\pm$0.08 & 210$\pm$58  & \hspace{-0.2cm}0.97 & H \\
2174 & 57 &  5 & 0.43 & 0.6798 &  3 & 22.08 & $-$0.49 & 0.04 & 0.46 & 43.07 & $-$21.00$\pm$0.09 & 175$\pm$24  & \hspace{-0.2cm}1.21 & H \\
2246 & 45 &  0 & 0.60 & 0.6514 &  5 & 21.52 & $-$0.58 & 0.04 & 0.37 & 42.96 & $-$21.27$\pm$0.09 & 183$\pm$29  & \hspace{-0.2cm}1.03 & H \\
2328 & 68 & 10 & 0.50 & 0.3956 &  8 & 22.44 & $+$0.26 & 0.06 & 0.61 & 41.66 & $-$20.16$\pm$0.11 & 150$\pm$18  & \hspace{-0.2cm}0.78 & L \\
2341 & 32 &  5 & 0.35 & 0.7611 &  5 & 22.09 & $-$0.38 & 0.04 & 0.31 & 43.37 & $-$21.25$\pm$0.09 & 279$\pm$107 & \hspace{-0.2cm}1.07 & H \\
2353 & 41 &  9 & 0.50 & 0.7773 &  3 & 22.05 & $-$0.29 & 0.04 & 0.35 & 43.43 & $-$21.48$\pm$0.09 & 209$\pm$44  & \hspace{-0.2cm}1.21 & H \\
2397 & 40 & 25 & 0.80 & 0.4519 &  5 & 21.98 & $+$0.42 & 0.06 & 0.34 & 42.00 & $-$20.84$\pm$0.08 &  98$\pm$22  & \hspace{-0.2cm}0.71 & H \\
2484 & 52 & 15 & 0.52 & 0.6535 &  5 & 21.78 & $-$0.58 & 0.04 & 0.41 & 42.97 & $-$21.07$\pm$0.09 & 169$\pm$36  & \hspace{-0.2cm}1.10 & H \\
2533 & 65 & 34 & 0.90 & 0.3150 &  5 & 20.42 & $+$0.04 & 0.06 & 0.56 & 41.08 & $-$21.32$\pm$0.09 & 230$\pm$16  & \hspace{-0.2cm}0.99 & H \\
2572 & 56 &  8 & 0.38 & 0.4491 &  5 & 23.00 & $+$0.41 & 0.06 & 0.45 & 41.98 & $-$19.90$\pm$0.10 &  74$\pm$19  & \hspace{-0.2cm}0.86 & H \\
2574 & 68 & 14 & 0.45 & 0.6802 &  5 & 22.98 & $-$0.53 & 0.04 & 0.61 & 43.07 & $-$20.22$\pm$0.13 & 162$\pm$29  & \hspace{-0.2cm}0.95 & L \\
2783 & 42 &  3 & 0.52 & 0.3143 &  5 & 21.57 & $+$0.04 & 0.06 & 0.35 & 41.07 & $-$19.95$\pm$0.08 &  84$\pm$17  & \hspace{-0.2cm}0.83 & L \\
2800 & 47 &  4 & 0.61 & 0.6290 &  8 & 22.23 & $-$0.64 & 0.04 & 0.38 & 42.87 & $-$20.42$\pm$0.09 & 143$\pm$27  & \hspace{-0.2cm}0.86 & H \\
2822 & 52 &  5 & 0.35 & 0.5871 &  8 & 21.80 & $-$0.71 & 0.04 & 0.41 & 42.68 & $-$20.62$\pm$0.09 & 107$\pm$21  & \hspace{-0.2cm}0.85 & L \\
2946 & 56 & 15 & 0.28 & 0.7437 & 10 & 22.99 & $-$0.42 & 0.04 & 0.45 & 43.31 & $-$20.38$\pm$0.13 & 108$\pm$34  & \hspace{-0.2cm}0.35 & L \\
2958 & 55 & 15 & 0.82 & 0.3139 &  5 & 21.48 & $+$0.04 & 0.06 & 0.44 & 41.07 & $-$20.13$\pm$0.08 & 141$\pm$13  & \hspace{-0.2cm}0.73 & H \\
3071 & 38 & 35 & 0.55 & 0.0939 &  5 & 19.53 & $+$0.34 & 0.08 & 0.34 & 38.17 & $-$19.39$\pm$0.08 & 176$\pm$20  & \hspace{-0.2cm}0.83 & H \\
3108 & 61 &  3 & 0.62 & 0.4741 &  5 & 23.33 & $+$0.46 & 0.06 & 0.50 & 42.12 & $-$19.81$\pm$0.11 & 104$\pm$26  & \hspace{-0.2cm}1.01 & H \\
3131 & 54 &  3 & 0.37 & 0.7723 &  5 & 22.47 & $-$0.36 & 0.04 & 0.43 & 43.41 & $-$21.05$\pm$0.12 & 166$\pm$26  & \hspace{-0.2cm}0.92 & L \\
3578 & 61 &  3 & 0.56 & 0.7718 &  5 & 22.68 & $-$0.36 & 0.04 & 0.50 & 43.41 & $-$20.91$\pm$0.13 & 154$\pm$37  & \hspace{-0.2cm}1.06 & H \\
3704 & 80 &  6 & 0.48 & 0.4082 &  3 & 23.92 & $+$0.42 & 0.06 & 0.96 & 41.74 & $-$19.27$\pm$0.16 &  40$\pm$20  & \hspace{-0.2cm}1.27 & L \\
3730 & 68 &  2 & 0.45 & 0.9593 &  5 & 22.97 & $-$0.95 & 0.04 & 0.61 & 43.99 & $-$20.72$\pm$0.15 & 156$\pm$41  & \hspace{-0.2cm}0.73 & H \\
3921 & 59 & 12 & 1.42 & 0.2251 &  3 & 19.90 & $+$0.82 & 0.08 & 0.48 & 40.25 & $-$21.73$\pm$0.11 & 245$\pm$19  & \hspace{-0.2cm}1.01 & H \\
4113 & 69 &  2 & 0.48 & 0.3951 &  1 & 23.04 & $+$0.46 & 0.06 & 0.63 & 41.65 & $-$19.77$\pm$0.12 & 249$\pm$53  & \hspace{-0.2cm}1.12 & L \\
4371 & 25 & 39 & 0.30 & 0.4605 &  3 & 23.13 & $+$0.52 & 0.06 & 0.30 & 42.05 & $-$19.80$\pm$0.09 & 361$\pm$330 & \hspace{-0.2cm}1.16 & L \\
4376 & 71 &  1 & 0.42 & 0.3961 &  8 & 23.72 & $+$0.27 & 0.06 & 0.67 & 41.66 & $-$18.94$\pm$0.11 &  74$\pm$18  & \hspace{-0.2cm}0.62 & L \\
4402 & 75 &  9 & 1.57 & 0.1138 &  3 & 20.24 & $+$0.47 & 0.08 & 0.78 & 38.62 & $-$19.71$\pm$0.09 &  95$\pm$3   & \hspace{-0.2cm}0.75 & H  \\
4465 & 61 &  9 & 0.40 & 0.6117 &  5 & 21.56 & $-$0.65 & 0.04 & 0.50 & 42.79 & $-$21.12$\pm$0.10 & 170$\pm$26  & \hspace{-0.2cm}1.04 & L \\
4498 & 40 & 40 & 0.33 & 0.7827 &  8 & 22.60 & $-$0.38 & 0.04 & 0.34 & 43.45 & $-$20.85$\pm$0.11 &  95$\pm$58  & \hspace{-0.2cm}0.82 & L \\
4657 & 37 &  5 & 0.70 & 0.2248 &  5 & 21.79 & $+$0.73 & 0.08 & 0.33 & 40.24 & $-$19.58$\pm$0.08 & 212$\pm$14  & \hspace{-0.2cm}0.91 & H \\
4730 & 32 &  3 & 0.90 & 0.7820 &  1 & 20.94 & $-$0.23 & 0.04 & 0.31 & 43.44 & $-$22.63$\pm$0.08 & 430$\pm$84  & \hspace{-0.2cm}1.41 & L \\
4806 & 65 &  4 & 0.70 & 0.2214 &  5 & 22.07 & $+$0.72 & 0.08 & 0.56 & 40.21 & $-$19.50$\pm$0.08 & 209$\pm$17  & \hspace{-0.2cm}0.72 & H \\
4922 & 64 & 13 & 0.71 & 0.9731 & 10 & 21.77 & $-$1.06 & 0.04 & 0.54 & 44.03 & $-$21.78$\pm$0.11 & 156$\pm$27  & \hspace{-0.2cm}0.22 & H \\
5022 & 73 & 10 & 0.61 & 0.3385 & 10 & 22.94 & $-$0.04 & 0.06 & 0.72 & 41.26 & $-$19.06$\pm$0.14 &  75$\pm$6   & \hspace{-0.2cm}0.51 & L  \\
5140 & 50 &  5 & 0.60 & 0.2738 &  5 & 22.94 & $-$0.08 & 0.06 & 0.40 & 40.73 & $-$18.16$\pm$0.10 & 125$\pm$39  & \hspace{-0.2cm}0.85 & L \\
5286 & 65 &  1 & 0.34 & 0.3337 &  8 & 23.88 & $+$0.06 & 0.06 & 0.56 & 41.22 & $-$18.02$\pm$0.18 &  23$\pm$7   & \hspace{-0.2cm}0.64 & L  \\
5317 & 53 &  5 & 0.59 & 0.9745 &  5 & 21.39 & $-$0.92 & 0.04 & 0.42 & 44.03 & $-$22.18$\pm$0.09 & 236$\pm$36  & \hspace{-0.2cm}0.71 & L \\ \\
\hline
\end{tabular}
\end{table*}

\begin{table*}
\addtocounter{table}{-1}
\centering
\caption{--- {\it continued}.}
\small
\begin{tabular}{cccccccccccccrc}
\hline \\
ID & \hspace{-0.2cm} $i$ & $\delta$ & $r_{\rm d}$ & $z$ & $T$ & $X$ & $k_B$ & \hspace{-0.2cm}$A^g_X$ & \hspace{-0.2cm}$A^i_B$ & 
\hspace{-0.2cm}$D_\Lambda$ & $M_B$  & $V_{\rm max}\hspace{6.5pt} $ & \hspace{-0.2cm}$B$$-$$R$ & note \\ 
 & \hspace{-0.2cm} [deg] &  [deg] & [\ $''$] & & & [mag] & [mag] & \hspace{-0.2cm}[mag]  & \hspace{-0.2cm}[mag] & \hspace{-0.2cm}[mag] & 
[mag] & [km/s]\hspace{0.7pt} & \hspace{-0.2cm}[mag] & \\  \\
\hline \\
5335 & 30 & 10 & 0.29 & 0.7726 &  5 & 22.53 & $-$0.36 & 0.04 & 0.31 & 43.41 & $-$20.86$\pm$0.10 & 220$\pm$135 & \hspace{-0.2cm}0.92 & H \\
5361 & 46 &  5 & 0.19 & 0.3339 &  5 & 22.45 & $+$0.10 & 0.06 & 0.37 & 41.23 & $-$19.32$\pm$0.09 &  75$\pm$19  & \hspace{-0.2cm}1.05 & L \\
5515 & 49 & 35 & 0.39 & 0.8934 &  3 & 21.33 & $-$1.03 & 0.04 & 0.39 & 43.80 & $-$21.86$\pm$0.10 & 269$\pm$87  & \hspace{-0.2cm}0.81 & H \\
5565 & 63 &  8 & 0.43 & 0.2285 &  8 & 23.55 & $+$0.68 & 0.08 & 0.53 & 40.28 & $-$18.01$\pm$0.11 &  41$\pm$10  & \hspace{-0.2cm}0.25 & L \\
6125 & 57 &  5 & 0.22 & 0.4495 &  5 & 22.78 & $+$0.41 & 0.06 & 0.46 & 41.99 & $-$20.14$\pm$0.08 & 152$\pm$18  & \hspace{-0.2cm}1.52 & H \\
6253 & 76 &  7 & 0.61 & 0.3453 &  5 & 23.75 & $+$0.14 & 0.06 & 0.81 & 41.31 & $-$18.58$\pm$0.19 &  95$\pm$19  & \hspace{-0.2cm}0.69 & H \\
6406 & 45 &  0 & 0.60 & 0.8451 &  3 & 21.65 & $-$0.16 & 0.04 & 0.37 & 43.65 & $-$22.25$\pm$0.08 & 250$\pm$36  & \hspace{-0.2cm}1.15 & H \\
6452 & 45 & 60 & 0.24 & 0.3359 &  5 & 23.15 & $+$0.11 & 0.06 & 0.37 & 41.24 & $-$18.63$\pm$0.11 & 109$\pm$52  & \hspace{-0.2cm}0.78 & L \\
6568 & 53 & 15 & 0.46 & 0.4597 &  5 & 22.85 & $+$0.43 & 0.06 & 0.42 & 42.04 & $-$20.11$\pm$0.09 & 144$\pm$30  & \hspace{-0.2cm}1.06 & L \\
6585 & 71 &  8 & 1.45 & 0.3357 &  5 & 20.26 & $+$0.11 & 0.06 & 0.67 & 41.24 & $-$21.83$\pm$0.08 & 295$\pm$7   & \hspace{-0.2cm}1.46 & H \\
6657 & 28 &  4 & 0.50 & 0.3343 &  3 & 20.99 & $+$0.18 & 0.06 & 0.30 & 41.23 & $-$20.78$\pm$0.07 & 254$\pm$54  & \hspace{-0.2cm}1.26 & H \\
6743 & 39 & 15 & 0.33 & 0.7320 &  8 & 22.73 & $-$0.47 & 0.04 & 0.34 & 43.27 & $-$20.46$\pm$0.10 & 118$\pm$48  & \hspace{-0.2cm}0.77 & H \\
6921 & 47 & 13 & 0.35 & 0.4538 &  5 & 23.47 & $+$0.42 & 0.06 & 0.38 & 42.01 & $-$19.40$\pm$0.12 & 111$\pm$46  & \hspace{-0.2cm}0.86 & H \\
7298 & 33 & 15 & 0.50 & 0.3902 &  5 & 21.48 & $+$0.29 & 0.06 & 0.32 & 41.62 & $-$20.81$\pm$0.09 & 115$\pm$56  & \hspace{-0.2cm}0.98 & H \\
7429 & 55 & 15 & 0.40 & 0.3370 &  5 & 21.68 & $+$0.11 & 0.06 & 0.44 & 41.25 & $-$20.19$\pm$0.09 &  90$\pm$17  & \hspace{-0.2cm}0.73 & L \\
7526 & 59 &  5 & 0.77 & 0.3589 &  5 & 20.86 & $+$0.19 & 0.06 & 0.48 & 41.41 & $-$21.28$\pm$0.07 & 181$\pm$7   & \hspace{-0.2cm}0.93 & H \\
7597 & 68 & 13 & 0.26 & 0.4096 &  5 & 24.24 & $+$0.34 & 0.06 & 0.61 & 41.75 & $-$18.52$\pm$0.21 &  79$\pm$34  & \hspace{-0.2cm}0.87 & L \\
7725 & 62 & 15 & 0.49 & 0.5504 &  8 & 22.18 & $-$0.77 & 0.04 & 0.52 & 42.51 & $-$20.12$\pm$0.12 &  84$\pm$38  & \hspace{-0.2cm}0.86 & L \\
7733 & 51 & 10 & 0.68 & 0.4471 &  5 & 22.79 & $+$0.41 & 0.06 & 0.41 & 41.97 & $-$20.06$\pm$0.09 &  89$\pm$17  & \hspace{-0.2cm}0.75 & H \\
7856 & 54 &  9 & 0.40 & 0.5146 &  5 & 22.63 & $+$0.52 & 0.06 & 0.43 & 42.32 & $-$20.69$\pm$0.09 & 273$\pm$35  & \hspace{-0.2cm}1.23 & L \\
7866 & 53 & 27 & 0.30 & 0.2240 &  3 & 23.11 & $+$0.82 & 0.08 & 0.42 & 40.23 & $-$18.44$\pm$0.10 &  50$\pm$45  & \hspace{-0.2cm}0.97 & L \\
8034 & 42 &  0 & 0.22 & 0.7317 &  5 & 22.34 & $-$0.43 & 0.04 & 0.35 & 43.27 & $-$20.89$\pm$0.11 & 173$\pm$49  & \hspace{-0.2cm}0.90 & H \\
8190 & 45 &  2 & 0.38 & 0.3947 &  5 & 21.84 & $+$0.31 & 0.06 & 0.37 & 41.65 & $-$20.55$\pm$0.08 & 113$\pm$12  & \hspace{-0.2cm}0.76 & H \\
8360 & 63 & 27 & 0.52 & 0.7034 & 10 & 22.58 & $-$0.48 & 0.04 & 0.53 & 43.16 & $-$20.67$\pm$0.16 & 120$\pm$62  & \hspace{-0.2cm}0.46 & L \\
8526 & 71 & 12 & 0.53 & 0.6095 &  5 & 21.90 & $-$0.66 & 0.04 & 0.67 & 42.78 & $-$20.94$\pm$0.14 & 135$\pm$10  & \hspace{-0.2cm}0.75 & L \\ \\
\hline \\
\end{tabular}
\end{table*}

\normalsize

$i$ --- Disk inclination derived by minimizing $\chi^2$ of a two--dimensional
exponential profile fit to the galaxy image in a coadded $I$-band frame 
with 0.49 arcsec FWHM (see Sect.~\ref{mags1}). 
In some cases, our initial constraint of $i\ge 40^\circ$ had to
be relaxed.

$\delta$ --- Absolute misalignment angle between the MOS slit and the 
apparent major axis. Our initial constraint of $\delta \le 15^\circ$ could
not always be met during the construction of the setups.

$r_{\rm d}$ --- Apparent disk scale length in arcseconds, 
derived via the same fits as
$i$ and $\delta$. While tests confirmed that $r_{\rm d}$ is the least 
critical input parameter in the
$V_{\rm max}$ derivation process, it is probably the one which is 
affected the strongest by the limitations of the ground--based imaging.
Hence, we recommend not to use these values for applications like
the Fundamental Plane of spiral galaxies.

$z$ --- Spectroscopic redshift.

$T$ --- SED type in the de Vaucouleurs scheme (see Sect.~\ref{class} for a
description of the classification criteria). $T=1$ accounts to Hubble type Sa, 
$T=3$ to Sb, $T=5$ to Sc, $T=8$ to Sdm and $T=10$ to Im.

$X$ --- Total apparent magnitude derived with the {\it Mag\_auto}
algorithm of the Source Extractor package
(Bertin \& Arnouts \cite{BA96}) on the coadded FDF frames. Depending on
the galaxies redshift, the filter $X$ was chosen to best match the rest--frame
$B$-band. For the least redshifted spirals with $z<0.25$, this was $X=B$,
for $0.25 \le z < 0.55$ we chose $X=g$, for $0.55 \le z < 0.85$ we 
set $X=R$, and for the highest redshifts in our sample with $z\ge0.85$
we used $X=I$.

$k_B$ --- K--correction for the transformation from the respective filter
$X$ (see above) to rest--frame $B$ as computed via synthetic photometry,
see Sect.~\ref{mags2} for details.

$A^g_X$ --- Galactic absorption in the respective filter $X$ (see above)
as given in Heidt et al. (\cite{Hei03}).

$A^i_B$ --- Intrinsic inclination--dependent dust absorption in rest--frame
$B$ following
Tully \& Fouqu\'e (\cite{TF85}) with the convention of a non--negligible  
extinction for face--on disks, i.e. $A^i_B=0.27^m$ at $i=0^\circ$.

$D_\Lambda$ --- Distance modulus in concordance cosmology with
$\Omega_{\rm m}$ = 0.3, $\Omega_\Lambda$ = 0.7 and 
$H_0$ = 70\,km\,s$^{-1}$\,Mpc$^{-1}$.

$M_B$ --- Absolute $B$-band magnitudes computed via 
$M_B=X-k_B-A^g_X-A^i_B-D_\Lambda$.
The errors include the uncertainties
in $X$, $k_B$ and $A^i_B$, see Sect.~\ref{mags2} for details.

$V_{\rm max}$ --- Intrinsic maximum rotation velocity derived via
synthetic velocity fields (see Sect.~\ref{vmax}) assuming a
linear rise of the rotation velocity at small galactocentric radii
and a flat RC at large radii (``rise--turnover--flat'' shape).
The errors on $V_{\rm max}$ were computed according to Eq.~\ref{verr}.

$B$$-$$R$ --- Rest--frame color index, corrected for intrinsic absorption.
For galaxies with $z<0.25$, apparent $R$ magnitudes were transformed to
rest--frame $R$, while apparent $I$ magnitudes were transformed to
rest--frame $R$ for all other objects, see Sect.~\ref{pairs} for details.

note --- Label indicating the RC quality. High quality curves (``H'')
extend well out to the region of constant rotation velocity at large radii.
RCs of low quality  (``L'') have a smaller radial extent and partly 
feature signatures of moderate kinematic perturbations like waves or
asymmetries.


\section{\label{tfr}The distant {$\mathbf B$}-band Tully--Fisher relation}

To be able to derive the spectrophotometric and/or kinematic evolution
of spirals at intermediate redshifts, it is crucial to carefully choose
a data set of spirals at low $z$ that can be used as reference.
Consistency between distant and local data set in terms of the intrinsic 
absorption correction is one of the key issues.
As stated in the introduction, a large number of local samples have been
constructed during the last decade, with kinematic data based on radio
observations or optical spectra. For our purposes, the term ``local'' refers to
redshifts below $z\le0.05$ corresponding to systematic velocities 
$V_{\rm sys} \le 15000$\,km/s. 
In Ziegler et al. (\cite{Zie02}), we selected a sample 
by Haynes et al. (\cite{Hay99}) as a local reference. Here, we will
use the data set of Pierce \& Tully (\cite{PT92}, PT92 hereafter) instead. 
Our initial choice was basically motivated by the very good statistics of
the Haynes et al. sample which comprises approx. 1200 spirals mainly of
type Sc, which is also the most frequent SED type in our distant sample. 

On the other hand, using the PT92 data makes
our results directly comparable to the studies of
e.g., Vogt  (\cite{Vog01}) and Milvang-Jensen et al. (\cite{Mil03}). 
Moreover, the TFR of PT92 benefits from the inclusion of spirals with
Cepheid--calibrated distances.
As for our sample, the photometry has been corrected for dust--reddening via
the Tully \& Fouqu\'e approach.
A difference that has to be accounted for is the convention for 
face--on extinction (see Sect.~\ref{mags2}) used by PT92, i.e., an offset of 
$\Delta M_B=0.27^m$ has to be applied. 
This way, the local $B$-band TFR transforms into
\begin{equation}
\label{pt92tfr}M_B=-7.48\log V_{\rm max}-3.52
\end{equation}
with an observed scatter of $\sigma_B=0.41^m$. This is in good agreement with
a bisector fit to the Haynes et al. sample restricted to 1097 galaxies with 
H\,{\small I} profiles classified as good quality by the authors: 
\begin{equation}
M_B=-7.85\log V_{\rm max}-2.78.
\end{equation}
Here, we transformed apparent $I$-band magnitudes into $B$ via
colors from Frei \& Gunn (\cite{FG94}). The two TFRs are perfectly consistent
in the low--mass regime and show only a small offset of $\Delta M\approx0.18^m$
at $V_{\rm max}=300$\,km/s.

\begin{figure}[t]
\resizebox{\hsize}{!}{\includegraphics{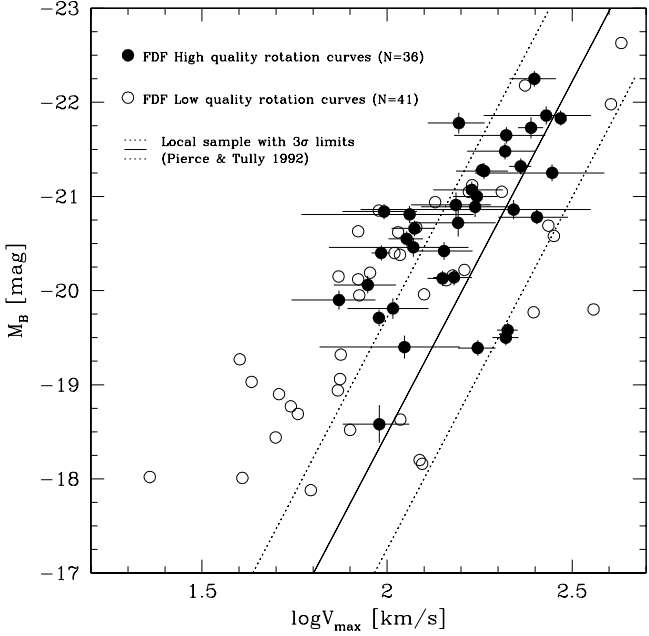}}
\caption{\label{tfrb}
FORS Deep Field sample of spirals in the range $0.1 \le z \le 1.0$ in comparison 
to the local TFR  by Pierce \& Tully (\cite{PT92}, dotted lines give 3\,$\sigma$
limits). The distant sample is
subdivided according to rotation curve quality: High quality curves (solid
symbols) extend well out to the region of constant rotation velocity at large
radii.
}
\end{figure}

In Fig.~\ref{tfrb}, we show the TF diagram of the complete FDF spiral sample
along with the local relation from PT92. As in Sect.~\ref{vmax}, our sample is 
sub--divided according to the RC quality. For the high quality data, the curves
have a sufficient spatial extent to probe the region of constant rotation velocity
at large radii, thereby yielding robust values of $V_{\rm max}$.
In the case of the low quality data, on the other hand,
we cannot rule out the possibility that at least for a fraction of the
objects $V_{\rm max}$ is underestimated, i.e. 
offsets from the local TFR towards higher luminosities could be overestimated.
For this reason, we will use only the high quality data for the following analysis.
These spirals cover the ranges $-18.6^m \ge M_B \ge -22.2^m$
in absolute magnitude and 
$74\,{\rm km/s} \le \log V_{\rm max} \le 295\,{\rm km/s}$
in maximum rotation velocity.

\begin{figure}[t]
\resizebox{\hsize}{!}{\includegraphics{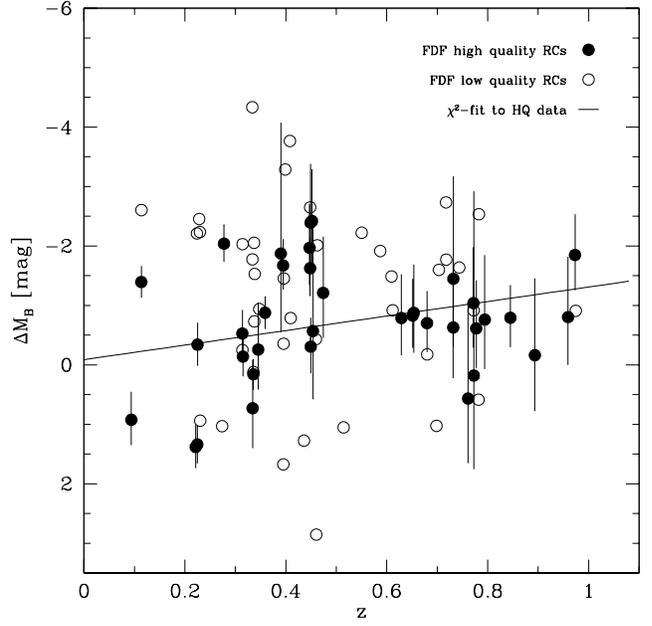}}
\caption{\label{tfres1}
Offsets of the FORS Deep Field sample from the local TFR
as given by Pierce \& Tully (\cite{PT92}) as a function of redshift. 
Filled symbols denote rotation curves which extend well out to the region of constant rotation 
velocity at large radii (labeled high quality data).
}
\end{figure}

For a given $V_{\rm max}$, the difference in luminosity of an FDF object
from the local PT92 fit is given by 
\begin{equation}
\label{respt}\Delta M_B=7.48 \log V_{\rm max}+3.52+M_B.
\end{equation}
These offsets are shown as a function of redshift in Fig.~\ref{tfres1}.
Although the scatter of the offsets is reduced from 
$\sigma_{\Delta M} \nobreak = \nobreak 1.32^m$ to 
$\sigma_{\Delta M} \nobreak = \nobreak 0.97^m$ when restricting the FDF sample to high quality RCs,
this is still over a factor of two larger than for the local data set.
We speculate that this partly is an effect of the observational limitations 
for distant spirals like, e.g., the low intrinsic spatial resolution, but also 
reflects a broader range of star formation
efficiencies than in the local universe. This interpretation is supported
by the smaller scatter of the offsets which originates from the uncertainty in
$V_{\rm max}$ and amounts to 
0.63\,mag 
for the HQ data.

A linear $\chi^2$-fit to the high quality data yields
\begin{equation}
\label{resz}\Delta M_B=-(1.22\pm0.56)\cdot z-(0.09\pm0.24).
\end{equation}
For this fit, the error was computed as
\begin{eqnarray}
\label{resz2}\sigma^2_{\Delta M_B} = \sigma^2_{M^{\rm FDF}_B} + 
7.48^2 \sigma^2_{\log V^{\rm FDF}_{\rm max}} \nonumber \\
+ \sigma^2_{M^{\rm PT}_B} +
7.48^2 \sigma^2_{\log V^{\rm PT}_{\rm max}},
\end{eqnarray} 
where the second and fourth term are the propagated errors from the 
uncertainties in $\log V_{\rm max}$ for the FDF spirals and the PT92 
sample, respectively.

As can be deduced from Eq.~\ref{resz}, we observe an increasing brightening
with rising look--back time. This is expected as an effect
of the younger stellar populations, i.e. a higher fraction of high--luminosity
stars than in the local universe.
Our result is in agreement with those of Barden et al. 
(\cite{Ba03}, $\Delta M_B=-1.1\pm0.5$ at 
$\langle z \rangle \approx 0.9$) and Milvang-Jensen (\cite{Mil03b}), 
who finds 
$\Delta M_B=-(0.9\pm0.3)\,\cdot\,z$, 
note that the latter correlation would be slightly steeper in the 
cosmology adopted here. 
A much smaller brightening of less than 0.2\,mag at $z\le1$
is found by Vogt (\cite{Vog01}).

\begin{figure}[t]
\resizebox{\hsize}{!}{\includegraphics{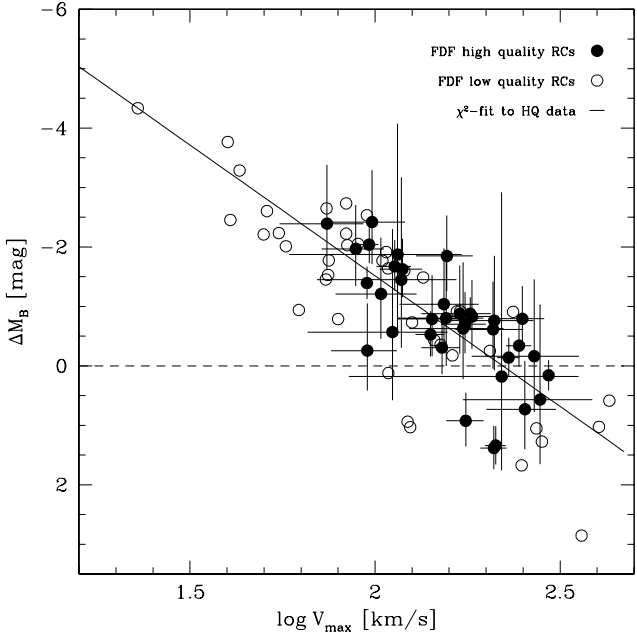}}
\caption{\label{tfres2}
Offsets of the FORS Deep Field sample from the local TFR
as given by Pierce \& Tully (\cite{PT92}) as a function of maximum rotation
velocity. We observe large overluminosities for objects with slow rotation
(i.e. low mass) and negligible offsets  for fast
rotators (i.e. high--mass spirals).
Filled symbols denote rotation curves which extend well out to the region of constant rotation 
velocity at large radii (labeled high quality data).
}
\end{figure}

Besides the dependency of $\Delta M_B$ on redshift, 
the comparison of our sample with PT92 in Fig.~\ref{tfrb}
indicates a correlation between the TF offsets and maximum rotation
velocity.
Even restricting to rotation curves which probe the ``flat'' region 
at large radii, a number of
distant spirals in the regime $V_{\rm max}\approx100$\,km/s are overluminous 
with $>$3\,$\sigma$ confidence, given the observed scatter of 0.41\,mag
for the local sample.
This can be seen in Fig.~\ref{tfres2}, where the offsets are plotted
against the logarithm of $V_{\rm max}$. 
A linear $\chi^2$-fit to the high quality subsample 
with an error estimation as defined in Eq.~\ref{resz2} yields
\begin{equation}
\label{resv}\Delta M_B=(4.40\pm0.69)\log V_{\rm max}-(10.31\pm1.55),
\end{equation} 
corresponding to a brightening by more than two magnitudes for the least
massive spirals in our sample and negligible offsets at 
$V_{\rm max}\approx 220$\,km/s,
which on the basis of Eq.~\ref{pt92tfr} is the typical rotation for local 
spirals of  luminosity 2-3$L^\ast$.

\section{\label{discuss}Discussion}

\begin{figure}[t]
\resizebox{\hsize}{!}{\includegraphics{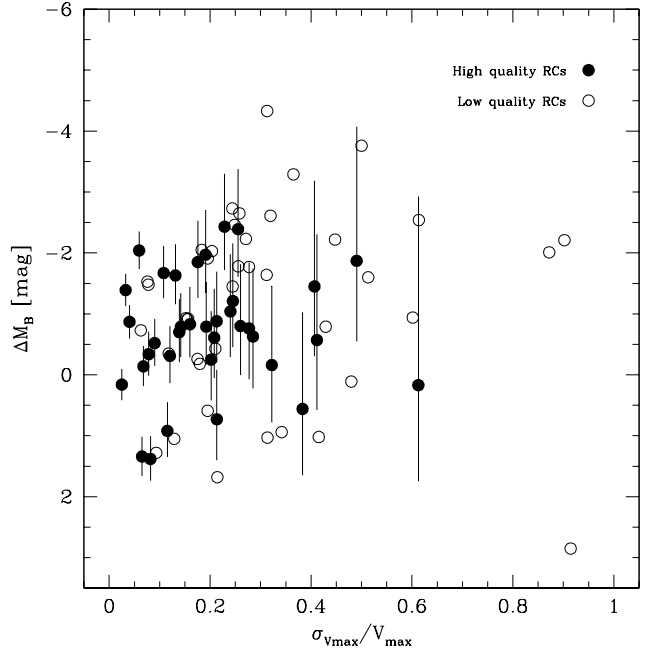}}
\caption{\label{resM}
Offsets of the FORS Deep Field sample from the local TFR
as given by Pierce \& Tully (\cite{PT92}), plotted against the relative 
errors in $V_{\rm max}$. The weak correlation between these two
parameters can attributed to the lower RC quality and the
larger relative errors of the RC extraction for slow rotators,
see text for details.  
Filled symbols denote rotation curves which extend well out to the region of constant rotation 
velocity at large radii.
}
\end{figure}

Before interpreting the possible physical implications of Eq.~\ref{resv},
we want to comment on  a potential correlation between the TF offsets 
and the errors in $V_{\rm max}$. In Fig.~\ref{resM}, the offsets are
plotted against the relative errors $\sigma_{V_{\rm max}}$/$V_{\rm max}$. 
Even if only the galaxies with high quality RCs are considered,
the overluminosities seem to be higher for objects with larger $V_{\rm max}$
errors.
We see two basic reasons for this slight dependency.

Firstly, the RC quality on the mean is lower for low--mass objects. In 
particular, the least massive FDF spirals all are classified as low
quality data. A lower rotation curve quality in turn leads to higher values of 
$\sigma_{\chi^2}$ (Eq.~\ref{verr}) and an increased total error of
the maximum rotation velocity.
And secondly, the errors of the Gaussian fits to the
emission lines which are performed in the process of the RC extraction
do not depend on the magnitude of the velocity shifts. On the other hand,
for a given radius and fit uncertainty, 
the relative errors on $V_{\rm rot}^{\rm obs}(r)$ are smaller
for fast rotators. Since the observed relative errors contribute to the
derived value of $\sigma_{\chi^2}$, this error
is on the mean larger for slow rotators. 

The combination of these two effects leads to larger relative errors
for small values of $V_{\rm max}$, i.e., for spirals with large TF offsets.

We will now consider potential biases due to galaxy-galaxy interactions
or sample incompleteness,
the impact of the intrinsic rotation curves shape and the question
of different conventions for the intrinsic absorption correction.

\subsection{\label{pairs}A bias due to environmental effects?}

To some extent, a correlation between rotation velocity and TF offsets is
to be expected from previous studies. Kannappan et al. (\cite{Kan02a}) 
found a color--residuals relation that reflects overluminosities of blue
spirals and argued that this could be attributed to enhanced
star formation. Since galaxies with blue colors, i.e. late types,
feature on the mean lower values of $V_{\rm max}$ than red, early--type
spirals (see Sect.~\ref{mass}), a correlation between colors and 
TF offsets should coincide with
a relation between the offsets and $V_{\rm max}$.  

To look into this effect, we computed the $B-R$ color index of our FDF galaxies in 
rest--frame. 
In contrast to the initial derivation of the absolute magnitudes,
we did not use total apparent magnitudes but 
brightnesses that were measured within apertures of two arcseconds diameter 
on coadded frames which were convolved to the same seeing
(see Heidt et al. \cite{Hei03} for details).
Similarly to the procedure for the  $B$-band luminosities, we
transformed the observed $R_{\rm FORS}$ into $R^{\rm rest}_{\rm Cousins}$ only
for low--redshift galaxies ($z<0.25$) and used the transformation
$I_{\rm FORS} \rightarrow R^{\rm rest}_{\rm Cousins}$ instead for all
objects at larger distances. The absorption coefficients with respect to rest--frame
$R$ were computed following Cardelli et al. (\cite{Ca89}) using again the
intrinsic absorption convention by Tully \& Fouqu\'e (\cite{TF85}).

\begin{figure}[t]
\resizebox{\hsize}{!}{\includegraphics{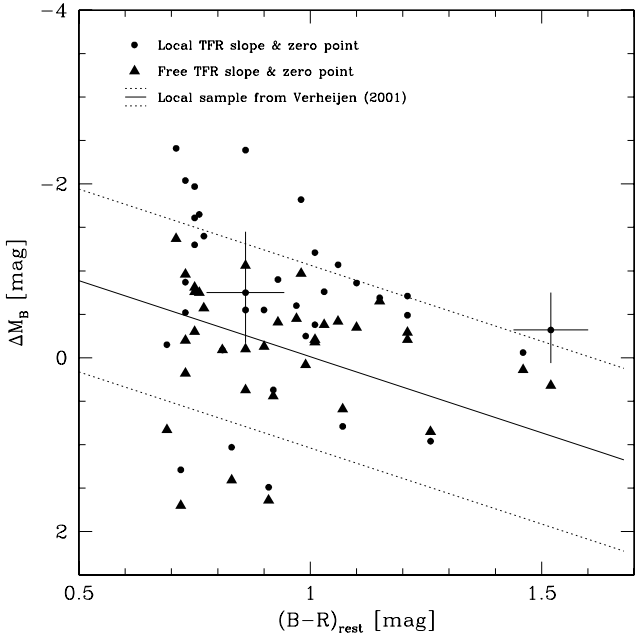}}
\caption{\label{tfres3}
Offsets of the high quality FORS Deep Field sample from the local TFR 
by Verheijen (2001, circles),
compared to the residuals of a free bisector fit to the high quality FDF data
(triangles), both as a function of rest--frame color.
The solid line represents the color--residual relation of the local
Verheijen sample, the dotted lines denote the 3\,$\sigma$ 
limits.
Typical error bars are shown for two FDF objects.
}
\end{figure}

For the purpose of the color--residual relation, we
compare our data to the local sample from Verheijen (2001).
We show the offsets with respect to the Verheijen TFR 
\begin{equation}
\label{tfrver} 
M_B=-8.1 \log V_{\rm max} - 2.07
\end{equation}
for our high quality
sample in Fig.~\ref{tfres3}. Only a weak dependence of the offsets on 
rest--frame color is observed. 
In contrast to the results for the local spirals,
offsets of the FDF spirals almost solely populate the regime of overluminosities 
with a median of $-0.7^m$ and 12 out of 36 galaxies feature offsets of more 
than one magnitude from the local TFR. Note that this is consistent with
the offsets from the PT92 relation which have a median of $-0.77^m$.

The situation is changed if one assumes that the TFR slope and zero point could
vary with cosmic time, i.e. if a free fit  is applied to
the distant sample. On the basis of the high quality data, a bootstrap bisector 
fit with 100 iterations yields
\begin{equation}
\label{tfrz051}M_B=-(4.66\pm0.67) \log V_{\rm max}-(10.43\pm1.50).
\end{equation}
To be consistent with the Verheijen sample, we performed the same fit
using the intrinsic absorption convention by Tully et al. (Eq.~\ref{tully1})
which yielded
\begin{equation}
\label{tfrz051b}M_B=-(5.22\pm0.69) \log V_{\rm max}-(9.08\pm1.56).
\end{equation}
New residuals using this relation for the FDF sample were derived analogous to
Eq.~\ref{respt} and are compared to the initial values in  Fig.~\ref{tfres3}.
The residuals computed via Eq.~\ref{tfrz051b} are mainly distributed 
within the range $-1\le \Delta M_B \le+1$ around zero, 
similar to the results 
given by Verheijen, though the scatter of the distant sample 
is larger and the correlation between color and residuals relatively weak. 
We conclude that the large offsets we find using Eq.~\ref{respt},
i.e. under the assumption of the local TFR slope and zero point,
can hardly be attributed to the color--residual relation
of local spirals.

Alternatively to an evolution of the TFR with look--back time,
one possible interpretation could be that a fraction of the distant
spirals are subject to
galaxy--galaxy interactions, which could result in TF offsets of 
up to several magnitudes as stated by
Kannappan et al. (\cite{Kan02b}). To perform a search for galaxy pair
candidates within our TF sample, we combined our complete data set  
(i.e. all galaxies with derived redshifts) 
with the lower resolution spectra from the FDF high--z campaign
(Noll et al. \cite{No03}), yielding a total of 267 galaxies at $z<1$. 
As upper limits on the three--dimensional separation of two  pair
candidates, we adopted the results of Lambas et al. (\cite{La02}).
Based on a data set of $\sim$\,10$^5$ objects 
from the 2dF Galaxy Redshift Survey,
the authors found that a
projected distance $D_{\rm proj} \le 100$\,kpc
and a relative radial velocity $\Delta V_{\rm sys} \le 250$\,km/s
are reliable upper limits to select galaxy pairs with enhanced
specific star formation.

\begin{figure}[t]
\resizebox{\hsize}{!}{\includegraphics{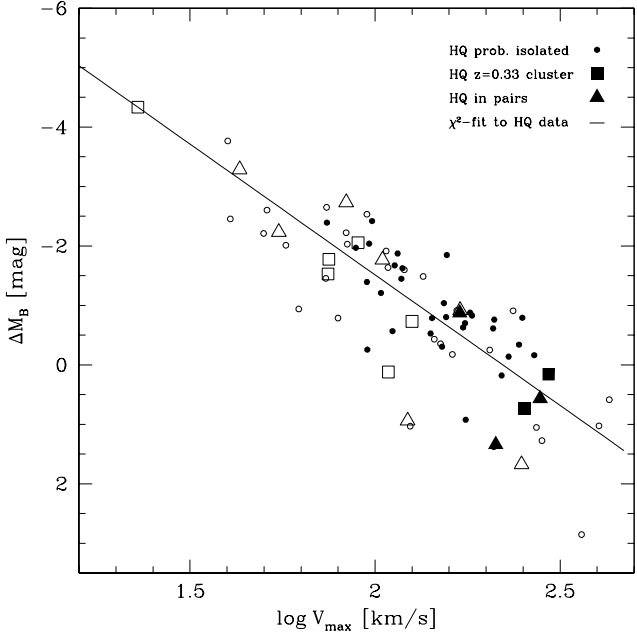}}
\caption{\label{tfres5}
Offsets of the FORS Deep Field spirals from the local TFR
as given by Pierce \& Tully (\cite{PT92}) as a function of maximum rotation
velocity. The FDF sample is sub--divided according to environment: 
Squares denote potential members
of a cluster at $z=0.33$, triangles are objects featuring neighboring
galaxies within $\Delta V_{\rm sys} \le 250$\, km/s and 
$D_{\rm proj} \le 100$\,kpc and circles are probably isolated galaxies.
Filled symbols denote objects with rotation curves that extend well into the
region of constant rotation velocity at large radii.
}
\end{figure}

12 spirals from the TF sample show spectroscopically confirmed neighbors
within these limits.
Of these galaxies, two are also possible members of the cluster at $z=0.33$
located in the southwestern corner of the FDF.
We show the TF offsets of these pair/cluster candidates in comparison 
to the rest of the sample in Fig.~\ref{tfres5}.
The two sub--samples seem to be similarly distributed.
This is also implied by a two--dimensional Kolmogorov-Smirnov test
which yields a propability of 0.31 that both subsamples are drawn from the
same distribution function.
We only observe a slight overrepresentation of pair/cluster candidates
towards low rotation velocities.  
These are all spirals with RCs classified as low quality data,
i.e. with relatively small radial extention and/or asymmetries.
In total, only 5 out of 18 (corresponding to 28\%) 
of the
pair/cluster candidates have high quality curves, whereas for the rest of
the sample this fraction accounts to 53\%. 
This is in agreement with
the results of Kannappan et al. (\cite{Kan02b}), who find that a fraction of
galaxies in close pairs show asymmetric or truncated RCs.
We therefore reject the idea that a significant fraction of the galaxies 
with high quality RCs are subject to interactions.

The nearest neighbor search cannot yield all pairs at $z<1$
within the FDF, since we do not have spectroscopic redshifts on all galaxies
in this regime (both the TF and the high--z study have a limit in apparent
brightness). However, the aim of this test was to clarify whether the
candidates which {\it are} identified differ from the rest of the sample.
Since the five pair/cluster candidates included in the analysis show only
moderate TF offsets, we conclude that it is unlikely that our high
quality subsample introduces a bias towards overluminosities
due to gravitational interactions.

\subsection{\label{incompl}Potential incompleteness effects}

In the following, we will adress the question of whether the 
deficiency of our sample of
objects with $M_B<-19$ and $\log V_{\rm max}>2.1$ (cf. Fig.~\ref{tfrb})
may be an observational bias. 

For a magnitude--limited TF sample, not all objects within the field--of--view
that are geometrically suitable enter the final data set. 
Towards the faint end  of the observed luminosity distribution, 
the sample therefore is incomplete.
At a fixed maximum rotation velocity, this may give rise to a bias against
low--luminosity spirals and, in turn, lead to an underestimated TFR slope as
pointed out by Giovanelli et al. (\cite{Gio97}).

Our apparent brightness limit of $R\le23^m$
corresponds to a limit in luminosity that is higher for larger redshifts. 
An impact of the incompleteness bias on our results should therefore coincide 
with a 
decrease of the TFR slope with increasing redshift.
However, if we split the complete FDF sample into sets for objects
with $z<0.45$ (41 galaxies) and 
$z>0.45$ (36 galaxies), the bootstrap fit slopes we find for the
two are $a=-3.75\pm0.44$ and $a=-3.77\pm0.59$, respectively.

Potentially, our sample may introduce a bias against fast rotators of 
low luminosity. This is because the spatially resolved emission lines of
large spirals (which on the average have higher $V_{\rm max}$ values) 
cover a larger CCD area and thus have
lower signal--to--noise ratios per pixel than small spirals. 
For a given total line flux, the data set could therefore
preferably contain slow rotators.

Since late--type spirals on average have stronger emission lines 
(and lower $V_{\rm max}$ values, see Sect.~\ref{mass}) than
early--type spirals, this effect would induce a type--dependency of the
TFR slope. For SED types $T\le3$, $T=5$ and $T\ge8$, we find 
slopes of $a=-3.74\pm1.40$, $a=-5.18\pm0.63$ and $-4.78\pm0.45$
(the large uncertainty
of the first of these fits shows the influence of low number statistics).
Although the 14 spirals of type Sb or earlier
have a flatter tilt than the two other sub--sets,
we do not find
a significant correlation between spectrophotometric type and TFR slope
within the derived errors of the bootstrap fits.

It is therefore unlikely that the shallower slope of the FDF sample with
respect to the local data is  
an observational effect.
We emphasize that, if the observed lack of distant spirals in the regime of
low luminosities and fast rotation was due to such a bias,
then the intrinsic scatter of the distant TF relation would have
to be increased by a significant factor compared to the local universe,
which we do not see.

\subsection{\label{urcs}Impact of the intrinsic RC shape}

As stated in Sect.~\ref{vfmod}, the rotation velocities derived via
the RC modelling are consistent for the ``rise--turnover--flat''
shape and the universal rotation curve from Persic \& Salucci (\cite{PS91}),
but slightly different if the URC as given in Persic et al. (\cite{PS96})
is used. However, if we restrict our sample to the high quality data, the
difference between $V_{\rm max}$ and $V_{\rm opt}$ is negligible
for spirals with $V_{\rm max}>150$\,km/s, 
and amounts to only 5\% in the median for the slow rotators.
In effect, the TF offsets 
would be altered towards lower luminosities 
by only $\sim$\,0.15$^m$
at $V_{\rm opt}=100$\,km/s if the URC96 was used alternatively.
None of our results would be affected significantly by such 
a small difference. 
For this reason, our analysis will be based on $V_{\rm max}$ 
throughout the section.

Another topic that has been referred to earlier are the slight differences
between $\chi^2$-fits and ``by--eye'' fits of the synthetic to the observed
rotation curves.
Since the former yielded systematically lower maximum rotation velocities
than the latter, the TF offsets 
would be {\it brighter}
by $\sim\nobreak0.25^m$ at $V_{\rm max}=100$\,km/s 
if the analysis was based on the $V_{\rm max}$ 
values from $\chi^2$-fit RC modelling as introduced in Sect.~\ref{vfmod}. 
Therefore, if one assumes a ``rise--turnover--flat'' shape of the intrinsic 
RC, our approach of a by--eye comparsion between synthetic and observed RC
yields conservative values of the TF offsets.

\subsection{Influence of the intrinsic absorption correction}

The convention we use for intrinsic absorption is purely inclination--dependent,
as the optical depth and fractional dust disk thickness are held fixed for
the entire sample.
More recently, Tully et al. (\cite{Tu98}) and Karachentsev et al. (\cite{Ka02})
have found evidence for an internal extinction law which also 
depends on rotation velocity. 
These observations favour
a higher amount of absorption in fast rotators than for spirals of low
$V_{\rm max}$. 
If one assumes the relation $2V_{\rm max}\approx W^i_R$ between 
maximum rotation velocity and H\,{\small I} profile linewidth
(see Tully \& Fouqu\'e \cite{TF85}), then equation 11 given in
Tully et al. transforms into
\begin{equation}
\label{tully1}A^{V,i}_B=2.75(\log V_{\rm max}-1.63)\log(a/b),
\end{equation}
where $a$ and $b$ are the apparent major and minor axes, respectively.
E.g., for a highly inclined disk with $i=80^\circ$, this yields 0.59$^m$ at
$V_{\rm max}=100$\,km/s and 1.35$^m$ at $V_{\rm max}=300$\,km/s,
whereas the initial Tully \& Fouqu\'e approach gives a value of
$A^i_B=0.96^m$ independent of the maximum rotation velocity. 

In effect,
the slope of any TFR would be steeper if the intrinsic absorption is
accounted for based on Eq.~\ref{tully1}. 
But, since this mass--depedency is linear in 
$\log V_{\rm max}/M_B$--space, the slope change would affect  
distant and local sample consistently. 
This can be quantitatively verified with Eqs.~\ref{tfrver} (the local Verheijen
TFR) and \ref{tfrz051b} (the FDF high quality sample) which are both derived
using the Tully et al. convention. Neither the median of the FDF spiral offsets
from the local TFR (0.7\,mag in the Tully  et al. convention vs. 0.77\,mag 
in the Tully \& Fouqu\'e convention) nor the evidence for a change in
the distant TFR slope ($>$3$\sigma$ confidence level in both conventions)
do significantly differ between the two approaches. 

Since our results are therefore independent of the intrinsic absorption
convention, we used the Tully \& Fouqu\'e approach throughout the paper
for the sake of direct comparability with the previous studies.

\subsection{\label{mass}A mass--dependent luminosity evolution?}

\begin{figure}[t]
\resizebox{\hsize}{!}{\includegraphics{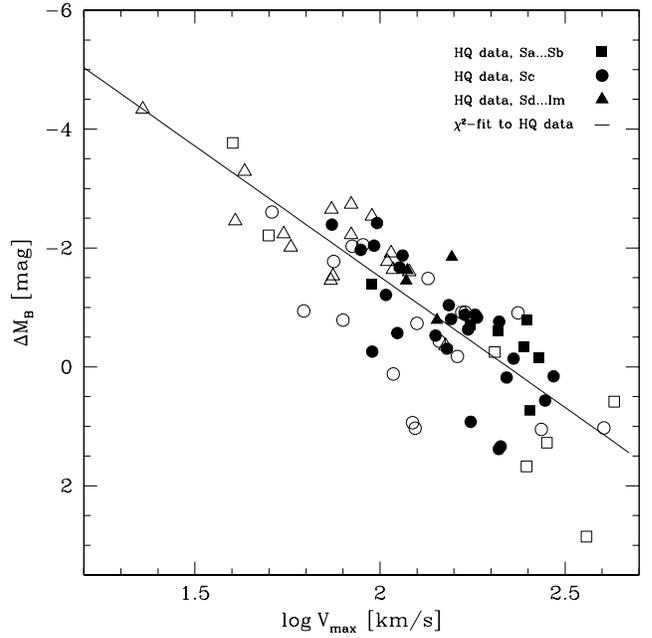}}
\caption{\label{tfres4}
Offsets of the FORS Deep Field sample from the local TFR
as given by Pierce \& Tully (\cite{PT92}) as a function of maximum rotation
velocity, sub--divided according to SED type into Sb or earlier (squares),
Sc (cicrles) and Sd or later (triangles), respectively. 
Filled symbols denote rotation curves which extend well out to the region of constant rotation 
velocity at large radii (labeled high quality data).
}
\end{figure}

It is well known that in the local universe, a dependency of 
$V_{\rm max}$ on galaxy type is observed. Blue, late--type spirals are on average
slower rotators 
than red, early--type spirals (e.g.  Rubin et al. \cite{Ru85}).
For our sample, sub--divided according to SED type, we find respective median
values of $\langle V_{\rm max} \rangle \approx 248$\,km/s for $T\le3$,
$\langle V_{\rm max} \rangle  \approx 145$\,km/s for $T=5$ and
$\langle V_{\rm max} \rangle  \approx 89$\,km/s
for $T\ge8$. 
Since we observe a correlation between the TF offsets and $V_{\rm max}$,
these $V_{\rm max}$ distributions imply a dependency of the mean luminosity
offsets on SED type as is illustrated in Fig.~\ref{tfres4}.

A consequence of this is a potential selection effect for small samples which
mainly comprise a certain sub--type. If a data set was biased 
towards late--type spirals due to the target selection on blue colors
(like, e.g., Simard \& Pritchet \cite{SP98}) or strong emission lines
(e.g.  Rix et al. \cite{Rix97}), a considerable evolution in luminosity
would be derived. On the other hand, if a data set preferably contains
early--type spirals, i.e. large disks (Vogt et al. \cite{Vog96}), 
only a modest luminosity offset from the local TFR will be observed.

A straightforward interpretation of the correlation between the luminosity
offsets and the maximum rotation velocity we find could be a change of the TFR
slope with look--back time. As given in Sect.~\ref{pairs}, a bootstrap 
bisector fit to the high quality FDF data yields a slope of $-4.66\pm0.67$,
corresponding to $>$99\% confidence for a shallower slope at intermediate
redshift with respect to the local PT92 sample. 
To verify this with a different fitting method,
we estimate the errors of the absolute magnitudes to be
\begin{equation}
\label{errfdf}\sigma^2_{\rm TFR} = \sigma^2_{M_B} + 
7.48^2 \sigma^2_{\log V_{\rm max}}.
\end{equation}
The first term on the right hand side corresponds to Eq.~\ref{errabs}
and the second term is the error propagation from the errors in $V_{\rm max}$
based on the PT92 slope. 
A linear $\chi^2$\,-fit to the high quality FDF then gives 
\begin{equation}
\label{tfrz052}M_B=-(3.55\pm0.41) \log V_{\rm max}-(12.84\pm0.92),
\end{equation}
i.e. the significance for a TFR slope change is even higher than on the basis 
of a bootstrap bisector fit.

As an explanation for the shallower slope at redshift 
$z \nobreak \approx \nobreak 0.5$,
we will now discuss two alternatives. Firstly, our result may point
to a luminosity evolution that depends on the maximum rotation velocity.
According to simulations by van den Bosch (\cite{vdB03}), the total mass
of a spiral galaxy within the virial radius can be estimated via
\begin{equation}
\label{vdB}M_{\rm vir} = 2.54\cdot10^{10} M_\odot
\Bigl(\frac{r_{\rm d}}{\rm kpc}\Bigr)\Bigl(\frac{V_{\rm max}}{100\,{\rm km/s}}\Bigr)^2, 
\end{equation} 
i.e. the correlation $\Delta M_B \propto \log V_{\rm max}$ implies
$\Delta M_B \propto \log M_{\rm vir}$. 
For the spirals with high quality RCs, covering the range 
$3.2 \nobreak \cdot \nobreak 10^{10} M_\odot \le M_{\rm vir} \le 1.6\cdot10^{12} M_\odot$
($\langle M_{\rm vir} \rangle = 2.3 \cdot10^{11} M_\odot$),
a linear $\chi^2$-fit yields  
\begin{equation}
\label{resmass}\Delta M_B=(1.14\pm0.27)\log \frac{M_{\rm vir}}{M_\odot}-(13.46\pm3.09).
\end{equation} 
Although the ground--based luminosity profile fits  may
place only upper limits on the scale lengths for some of the apparently
smallest  galaxies, 
we find evidence for a mass--dependent luminosity evolution which accounts to
$\sim$\,2$^m$ in rest--frame $B$\,
for the least massive objects and is negligible for high--mass spirals.
This implies that the redshift dependency we observe (Eq.~\ref{resz})
is most probably a lower limit, since low-mass spirals --- 
which show strong evolution in luminosity according to Eq.~\ref{resmass} ---
at higher redshifts will fall beyond our apparent brightness limit.
For the same reason, we cannot speculate on a potential evolution 
(more precisely, a decrease) of the mean galaxy masses with redshift. 

If one assumes that low--mass spirals have not undergone a
significant increase of their metallicities since $z  \nobreak \approx  \nobreak 0.5$,
our result is similar to that of Kobulnicky et al. (\cite{Ko03}),
who used the distant luminosity--metallicity relation and found 
$\Delta M_B$~$\approx$~$-$1\dots$-$2\,mag.
On the other hand, Vogt (\cite{Vog01}) does not find a TFR slope change
with a TF sample from the same survey (DEEP Groth Strip Survey).

Besides a mass--dependent luminosity evolution, a second possible explanation 
for the flatter tilt
of the distant TFR could be a strongly starforming galaxy population within our 
sample that contributes less to the local luminosity density,
i.e. either these galaxies could be overnumerous or overluminous at intermediate
redshift. In terms of the luminosity
function, this topic is well known as the ``faint blue galaxy excess''
(see Ellis \cite{Ell97} for an overview). Based on spectroscopic data from the
Canada-France-Redshift-Survey, Driver et al. (\cite{Dri96}) found that
dwarfs may have faded by over one magnitude between the regime $0.2<z<0.5$ 
and the local universe.
These observational findings could be understood in terms of 
a reionization era between redshifts $z \approx 6$ and  $z \approx 1$
which suppresses star formation in low--mass dark halos
($M_{\rm vir} \le 10^9$\,$M_\odot$, Babul \& Rees \cite{BR92}). 

However, 
only a small fraction of the high quality FDF sample could 
be members of such a blue dwarf population 
since the former covers the range  
$74\,{\rm km/s}~\le~\log V_{\rm max}~\le~295\,{\rm km/s}$ and 
the lower virial mass limit of $3\cdot10^{10} M_\odot$ 
also exceeds the dark halo mass range of dwarfs.
Furthermore, the star formation rates derived from [O{\small II}] 
equivalent widths fall in the
range 
$0.8\,M_\odot\,{\rm yr}^{-1}
\le {\rm SFR} \le 
18.3\,M_\odot\,{\rm yr}^{-1}$,
with a median of
$\sim$\,4.5\,$M_\odot\,{\rm yr}^{-1}$, 
typical for massive
local spirals (e.g. Kennicutt \cite{Ken83}).

If our result is interpreted in terms of a decreasing luminosity of
low--mass spirals over the past $\sim$\,5\,Gyrs (which accounts to the 
look--back time at $z\approx0.5$), 
it could be due to a mass--dependent evolution of the mass--to--light
ratio, or even a mass--age relation.
A direct comparison
to predictions of stellar population models is difficult since any evolution 
of the TFR introduces several competing effects.
On the one hand, younger stellar populations coincide with a decrease of
the mass--to--light ratio.
On the other hand, since less gas has been consumed via star formation,
the gas mass fraction most probably increases with look--back time,
thereby increasing the mass--to--light ratio.
Moreover, within the framework of hierarchical merging, disk sizes
should be smaller for a given maximum
rotation velocity and masses on the mean lower towards higher redshift 
(e.g.  Mao et al. \cite{MMW98}). Based on the rotation velocity--size relation, 
i.e. the correlation between $V_{\rm max}$ and $r_{\rm d}$, our sample 
shows some  evidence for slightly smaller disks at $z\approx0.5$
(see B\"ohm et al. \cite{Boe03}). However, due to the limitations of the
ground--based imaging, the significance of this evolution is low. 

In effect, a decrease of the disk sizes and an increase of the gas mass 
fraction would tend
to shift distant spirals to the low--luminosity side of the local TFR, whereas
a lower stellar mass--to--light ratio would result in a shift to the
high--luminosity side. A domination of the first two processes for fast
rotators could explain
why a fraction of the high--mass FDF spirals are underluminous in the
TF diagram.

As already stated, the correlation between luminosity evolution and redshift 
given in Eq.~\ref{resz} is probably only a conservative estimate.
Because the disk sizes and gas fractions may also evolve,
the pure luminosity evolution possibly is larger
than approx. one magnitude between $z=1$ and the local universe.
E.g., chemically consistent evolutionary synthesis models by
M\"oller et al. (\cite{Moe01}) yield a luminosity evolution of
$\sim$\,1.5\,mag in rest--frame $B$ for an Sc spiral over this
redshift range.

Although the mild evolution of the scale lengths that might be deduced 
from the FDF
sample is in compliance with the ``bottom--up'' scheme of structure growth
within the Cold Dark Matter hierarchical model, 
a luminosity evolution that
is larger for low--mass galaxies 
would be in contradiction to it. If our result is interpreted in terms of
younger ages (lower formation redshifts) for lower masses, the contradiction 
will  be even more obvious.
This is simply because small Dark Matter Halos should have formed earlier
than large ones, and thus the stellar populations in galaxies of low mass
should be older than those of high-mass systems.
Semi--analytic simulations which account for a mass--dependent evolution of 
the stellar populations also yield an increase of TFR slope with look--back
time (e.g.  Ferreras \& Silk \cite{FS01}).

On the other hand, it is well established and also valid for our sample that 
the colors of spirals tend to become redder with mass (beginning of this 
section) which is not reproduced by simulations within the hierarchical
scenario unless the spectrophotometric properties of local spirals
are used as a calibration
(e.g.  Bell et al. \cite{Be03}). 

We therefore conclude that, if the TFR slope of our distant sample 
is related to a mass--dependant luminosity evolution,
this would be at variance with the hierarchical merging scenario 
on small scales.


\section{\label{conclude}Conclusions}

Using imaging data and spectroscopy taken with the ESO Very Large Telescope, 
we have derived structural parameters 
and resolved rotation curves of a magnitude--limited sample of 77 
spiral galaxies in the FORS Deep
Field. The objects cover the redshift range $0.1 \le z \le 1.0$ and
comprise all types from Sa to very late--type.
Via a rotation curve modelling that takes into account geometric effects
as well as seeing and optical beam smearing, the maximum rotation velocities
have been derived and the distant $B$-band Tully--Fisher relation was constructed.

We find evidence for a luminosity evolution with look--back time
which amounts to a brightening of $\Delta M_B \approx -1^m$ at redshift unity.
Moreover, we observe a correlation between the luminosity evolution and the
total masses. The distant low--mass spirals are brighter by up to two magnitudes than 
their local counterparts, whereas the luminosity evolution of high--mass
systems is negligible. In effect, the slope of the Tully--Fisher relation
at intermediate redshift is shallower than for local samples.
This may partly be caused by a population of small, starforming
galaxies that contribute less to the luminosity density in the local 
universe. 
Nevertheless, the vast majority of our objects have virial masses too
large to be dwarf galaxies and show star formation rates typical
of normal spirals.
The flatter tilt we find for the distant Tully--Fisher relation 
is in contradiction to the predictions of recent
semi--analytic simulations.


\begin{acknowledgements}
We acknowledge the thorough and useful comments of the referee
and are grateful 
for the continuous support of our project by the PI of the FDF
consortium, Prof. I.~Ap\-pen\-zel\-ler.
We also thank ESO for the efficient support 
during the observations and 
Drs. 
B. Milvang-Jensen (MPE Garching) and M.~A.~W.~Ver\-hei\-jen (Potsdam) 
for helpful comments. 
Our work was funded by the Volkswagen Foundation (I/76\,520), 
the Deutsche Forschungsgemeinschaft (SFB375, SFB439)
and the German Federal Ministry for Education and Science
(ID--Nos. 05\,AV9MGA7, 05\,AV9WM1/2, 05\,AV9VOA).
\end{acknowledgements}


\end{document}